\documentclass[12pt]{spieman}
\usepackage{amsmath,amsfonts,amssymb}
\usepackage{graphicx}
\usepackage{setspace}
\usepackage{booktabs}
\usepackage{orcidlink}

\title{Multi-modal transmission forecasting for the Vera C. Rubin Observatory}

\author[a]{Peter Xiangyuan Ma\,\orcidlink{0000-0001-8975-3719}}
\author[a]{Joshua S. Bloom\,\orcidlink{0000-0002-7777-216X}}
\affil[a]{Department of Astronomy, University of California, Berkeley, 501 Campbell Hall, Berkeley, CA 94720, USA}

\begin{document}
\maketitle

\begin{abstract}
In the age of large automated surveys conducted at facilities like the NSF-DOE Vera C.\ Rubin Observatory, delivering precision science goals demands highly efficient scheduling strategies that adapt to changing environments on the summit. On site, a variety of telemetry streams can be used to inform a scheduler that is incentivized to find the best next pointing given atmospheric transmission. At Rubin, live upward-facing monitoring is provided by the Dutch Rubin Enhanced Atmospheric Monitor (DREAM) all-sky camera. Currently, dynamic scheduling uses short-term predictive algorithms, such as tracking cloud cross-correlations, to anticipate movement at short ($< 10$\,min) scheduling timescales. However, many of Rubin's core science objectives require survey strategies to commit to observations 20--40\,min in advance (pair mode), which remains out of reach because clouds at those timescales can arrive from beyond the horizon. Here, we report on a novel multimodal strategy that folds asynchronous data from both the DREAM camera and weather satellite imagery to forecast multi-horizon sky transmission at 90\,s, 5\,min, 20\,min, and 30\,min in advance. We demonstrate that at 20-minute timescales, our model reduces transmission error by 21\% and false-clear misses by 22\% versus Rubin's frozen-flow tracker, and reduces false-clear misses by 26\% versus persistence. We also demonstrate empirically, through ablations, that the DREAM all-sky camera anchors the short-horizon predictive power while satellite imagery provides long-horizon predictive power. We estimate the optimistic order-of-magnitude gain for LSST survey efficiency improvement to be between 0.1\% and 1\%. 
\end{abstract}

\keywords{transmission forecasting, all-sky camera, cloud forecasting, satellite imagery, dynamic scheduling, Vera C. Rubin Observatory, multimodal learning}

{\noindent \footnotesize\textbf{*} \linkable{peter\_ma@berkeley.edu} }

\begin{spacing}{2}

\section{Introduction}
\label{sect:intro}
The NSF-DOE Vera C.\ Rubin Observatory's Legacy Survey of Space and Time (LSST) is expected to image the entirety of the southern night sky every $\sim$3.5\,days for the next decade. This will deliver, at its maximum depth with coadded images, a limiting magnitude of $\sim$27.5~\cite{Ivezic2019}. With this capability, the collaboration will provide a treasure trove of new data that will revolutionize multiple fields, ranging from solar system science, mapping Galactic structure, and uncovering the transient sky to constraining the dark sector, including dark energy.

The success of the survey hinges on providing highly efficient and optimized observing strategies. This is because many science goals tolerate only a few percent in survey efficiency. Examples of such demanding targets include finding and tracking hazardous Near-Earth Objects~\cite{Jones2018}, Type~Ia supernova cosmology~\cite{Sanchez2022,Lochner2022}, determining microlensing caustic crossings, kilonova characterization in Targets of Opportunity, and more. The sensitivity of these programs to scheduling strategies is primarily due to the time domain component of their science. It is often not enough for a single observation to characterize these targets, and multiple visits of the same field during the same night are required. This presents itself as a challenging scheduling problem.

For example, one of the common scheduling patterns is the Wide-Fast-Deep (WFD) pair cadence: two mixed-filter visits $\sim$20--40\,min apart (median of 33\,min), which currently occupies slightly more than 80\% of survey time~\cite{PSTN055,LSSTObservingStrategy,Schwamb2023}. The challenge for the scheduler is that any collapse in the second observation can render the data unusable, for example for Solar System linking completeness~\cite{Jones2018}. This, had it been foreseen, could have been repurposed to deepen coadds or to serve alerts and other parts of the survey campaign, instead of being wasted.

The current deployed scheduler is responsive to change and is dynamic in incorporating telemetry and information live on the summit to arbitrage survey strategies. However, the least constrained piece of information is \textit{anticipating} the weather, particularly on medium ($\sim$20--40\,min) horizons. Currently, the all-sky camera puts cloud cover at $\sim$20--30\% of scheduled time at ground-based sites~\cite{Benn1998,Qian2024,Xu2023,Haslebacher2022}. At Cerro Pach\'on about 53\% of scheduled time is photometric (near-perfect transmission). Another $\sim$10\% is patchy or thin cloud: still usable if the telescope can point around the opaque sky rather than through it~\cite{LSST_OSS,Kerber2014,Kerber2016}. Traditionally, an observer during a patchy night can intuit by eye where one might successfully point for their next observation. A human observer using a mix of summit telemetry and live weather reporting can inform the operator well enough to ``dodge'' large clouds in the field of view of upcoming targets. An automated survey, like LSST, cannot~\cite{Tokovinin2023,Patat2011,Masciadri2013}. As such, this leaves room for a few percent level improvement in observing efficiencies without introducing any hardware changes at the summit.

\subsection{LSST Survey Strategy}
\label{sect:lsst}
Currently, the survey strategy uses the Feature-Based Scheduler (FBS) to allocate pointings. The FBS chooses from a variety of survey patterns, splitting its time among WFD ($\sim 80\%$ of total time), Deep Drilling Fields (DDFs; $\sim$7\%, intensive revisits of five pointings, one magnitude deeper than WFD), mini- and microsurveys (a few percent to $\sim$10\%: Galactic plane, North Ecliptic Spur, twilight), and Targets of Opportunity (ToOs) ($\sim$3\%)~\cite{Ivezic2019,Bianco2022,PSTN055,LSSTObservingStrategy}. These strategies sit in a priority tree: ToOs, coordinated synergies, DDFs, template building, intra-night triplets, pair visits, twilight, then a greedy single-visit filler~\cite{Naghib2019,Jones2021,Schwamb2023}. It ranks pointings from live telemetry: Differential Image Motion Monitor (DIMM) seeing, Dutch Rubin Enhanced Atmospheric Monitor (DREAM) transparency, Environmental Sensor Station (ESS) wind, and mount state, together with Sun/Moon geometry and a sky-brightness model~\cite{Stuik2024,Yoachim2016,SMTN002,TSTN035}.

At the moment, a standard visit takes $\sim$30\,s on sky and commits with two further visits sitting in the queue in a single batch. Thus at any given time, three pointings are in-flight with an assumed lookahead time of  $\sim$90\,s~\cite{Ivezic2019,Jones2021}. This demands dynamic forecasting with a horizon of a few minutes (1--3\,min). To address this, a new dynamic scheduler~\cite{Segev2026} uses a frozen flow model and cross-correlation of cloud maps to produce scheduler updates from data collected by an on-site all-sky camera (DREAM). 

However, there exist other survey strategies that are far more committing than the immediate-next-observation plan. For example, baseline WFD visits are typically obtained in mixed-filter pairs separated by $\sim$33\,min, so pair-mode is the default for the $\gtrsim$80\% of survey time spent on WFD~\cite{PSTN055,LSSTObservingStrategy,Schwamb2023}. A third visit in the same filter, 3--7\,h after the pair, is acquired for only $\sim$3\% of pairs ($\sim$4\% of all visits), about once every seven nights; three visits per night is not the default~\cite{PSTN055,LSSTObservingStrategy}. Typically, if a cloud ruins the second visit, the first visit is mostly wasted for its science case, so these modes require forecasting at least 20\,min ahead. This remains the most challenging domain, as clouds can form beyond the horizon and simply cannot be fully captured by local telemetry on the mountain alone. 

\subsection{DREAM Camera}
\label{sect:dream-intro}
One of the key pieces of telemetry is the DREAM all-sky camera on the summit. DREAM produces high-cadence ($\sim$30s) all-sky transparency on site, built from the Multi-site All-Sky CAmeRA (MASCARA) and bRing~\cite{Stuik2022,Stuik2024,Snellen2013,Talens2017,Stuik2017}. These fixed cameras are pointed at zenith and the four cardinal directions, covering $53^\circ\times74^\circ$ each at $\sim$1$'$\,pixel$^{-1}$, down to airmass $\sim$3 ($\sim$20$^\circ$ altitude)~\cite{Talens2017,Stuik2022}. DREAM photometers catalog stars with $V<8.4$ every 6.4\,s and emit nested Hierarchical Equal Area isoLatitude Pixelization (HEALPix) maps of gray cloud extinction $A$ matched in lockstep to the LSST visit cadence~\cite{Talens2018,Gorski2005,Stuik2024}. The intended use is live FBS input on non-photometric nights~\cite{Stuik2024,Naghib2019}.

\subsection{Dynamic Scheduling and Forecast Horizons}
To leverage the telemetry from DREAM, the dynamic scheduling introduced by Segev~\cite{Segev2026} showed that a responsive FBS, operating at short timescales, can recover substantial photon yield on patchy nights by tracking cloud motion to choose the next pointing: on representative nights, +66\% photons and +0.67\,mag single-visit depth versus the status quo baseline, recovering most of the clear-patch bound. This delivers substantial gains on short-term scheduling; however, it still leaves pair and triplet modes unprotected~\cite{LSSTObservingStrategy,PSTN055}. 

At its core, any algorithmic prediction method is limited by how fast clouds leave the scored sky. Above $30^\circ$ altitude, a deck $1$--$7$\,km above the $2.6$\,km summit spans about $10$--$25$\,km. At high-cloud speeds of $15$--$20$\,m\,s$^{-1}$ that field crosses in about $10$--$25$\,min; beyond that window the frozen-flow tracker has no remaining structure on the scored sky, and predictability falls off as the flow becomes chaotic~\cite{Segev2026}. The full camera reaches $\sim$20$^\circ$ altitude and, for the same decks, is about $5$--$40$\,km across. Pair-mode WFD at $20$--$40$\,min still sits at or past one crossing of the scored field. A greedy scheduler thus cannot protect the second visit of a $\sim$20--40\,min pair. Cloud avoidance at that horizon needs a full-sky transmission forecast tens of minutes ahead, including cloud that has not yet entered the camera frame.

\subsection{Classical Transmission Forecasting}
\label{sect:classical}
Forecasting, sometimes called nowcasting at short horizons, has been studied extensively. Classical models rely on the kinematics of historically observed data to forward-model trajectories of future events. Often, on short timescales, the zeroth-order solution, and sometimes a sufficient one, is to assume that nothing changes: Eulerian persistence from frame to frame~\cite{Germann2002,Seed2003}.  It has been shown that within short timescales this can be competitive; for example, some all-sky solar imagers used the same freeze as the intra-hour reference~\cite{Chow2011,Yang2019}. The competitiveness of these simplest possible solutions is that standing cloud decks and slow cirrus can often change little over tens of minutes. Thus a copy can often be predictive. The clear bottleneck here is that it encodes no motion; cloud that translates, grows, or enters from off-frame is a miss~\cite{Pulkkinen2019}.

The next upgrade to persistence is to estimate the motion successively through multiple images and then advect and move the clouds. In many cases, Lucas and Kanade's local optical flow (OF)~\cite{Lucas1981,Bouguet2000} can be used as the velocity vector for rain-rate nowcasting~\cite{Bowler2006} or for all-sky cloud tracking~\cite{Chow2011}. At its core, OF assumes that objects remain constant in brightness and, under that assumption, tracks pixels in time to forward-model the next step. Further advances include semi-Lagrangian backtraces~\cite{Staniforth1991}, a mix of the Eulerian scheme (stationary grids) and a fully Lagrangian scheme with moving grids. Semi-Lagrangian advection still uses a fixed Eulerian grid to store transmission values, but filling a mesh point follows a fluid parcel. Further methods include scale-cascade schemes such as S-PROG~\cite{Seed2003} and pysteps~\cite{Pulkkinen2019}, which keep the same Lagrangian skeleton while matching lifetime to spatial scale. However, these frozen-flow approximations cannot represent cell growth or decay, stacked layers at different speeds, or cloud that has not yet entered the frame. The velocity estimate is also noisy.

Later work combined standard optical-flow models with telemetry from physical wind measurements to reduce the noise in the gradient calculation. That kinematic line replaced the camera velocity with an independent meteorological wind. Cloud outside the camera at $t_{\mathrm{end}}$ nevertheless remains invisible at long enough horizons.

The current proposed dynamic scheduler for the FBS, detailed in~\cite{Segev2026}, implements a similar camera-inferred motion model. A frame-to-frame cross-correlation under a frozen-flow assumption uses the DREAM maps to determine the next pointing. Its primary goal is to produce a clean global cloud velocity vector from successive DREAM frames, then use it to keep the telescope inside the moving clear hole for the next $\sim$32\,s visit. The velocity estimate dithers the transmission map in various directions and magnitudes, computes a Pearson correlation with the true successive frame, and selects the perturbation that maximizes the correlation. The optimal perturbation is used to compute the global velocity vector. This has produced large immediate gains in survey efficiency~\cite{Segev2026}. 

\subsection{Deep-Learning-Based Forecasting}
\label{sect:datadriven}
Classical models' reliance on optical-flow or frozen-flow assumptions is a limitation of their design. Upgrading those models to full fluid simulations remains inaccessible at these nowcast latencies. There is therefore a growing interest in data-driven solutions, with the goal of learning that residual from a window of fields, without an explicit velocity, from the observed data itself.

One of the foundational architectures in this problem domain is the convolutional long short-term memory (ConvLSTM)~\cite{Shi2015} model. Naturally, one can imagine casting this as a sequence-modeling problem. However, a standard long short-term memory (LSTM) treats each timestep as a feature vector, which destroys spatial structure and is too expensive for a map. ConvLSTM replaces those gates with convolutions so the hidden and cell states remain spatial fields. Recurrence then acts as an implicit memory of local advection. The network can learn to move, grow, or fade structure that a frozen-flow tracker can only translate. Later, recurrent models added longer temporal memory (PredRNN)~\cite{Wang2017}, trajectory-gated motion (TrajGRU)~\cite{Shi2017}, or a partial differential equation (PDE)-constrained branch beside a residual cell (PhyDNet)~\cite{LeGuen2020}. 

A second family of models dropped this recurrence structure. Simpler video prediction (SimVP)~\cite{Gao2022} stacks the history on the channel axis and runs an encoder--translator--decoder: spatial mixing in the encoder and decoder, a residual convolutional neural network (CNN) translator at the bottleneck. The claim is that a feed-forward translator on stacked frames matches or beats recurrent video models without the cost of unrolling. Larger models then brought cuboid and axial attention (Earthformer, MetNet)~\cite{Gao2022Earthformer,Sonderby2020} and generative nowcasts (DGMR, NowcastNet)~\cite{Ravuri2021,Zhang2023}.

In this work, we choose to focus on the two architectural paradigms, ConvLSTM and SimVP, because other larger models require much longer data baselines, while the present work supplies only one year of training data.

\subsection{Multimodal Data from Satellites}
\label{sect:satellite-intro}
Since a key challenge in transmission forecasting is in predicting cloud movement from beyond the horizon, a clear path to resolving the problem is to simply measure clouds from beyond the horizon. This can be provided via satellite imagery. Recently, there has been a proliferation of publicly available geosynchronous weather-satellite data; live feeds of the region surrounding the summit can be retrieved. For example, an upwind view can be recovered from the Advanced Baseline Imager (ABI) on the National Oceanic and Atmospheric Administration (NOAA) Geostationary Operational Environmental Satellite (GOES-East; GOES-16, then GOES-19)~\cite{Schmit2017,NOAAGoes19}. These instruments provide night-capable infrared products at 10\,min / $\sim$2\,km (C13 brightness temperature, the ABI Cloud Height Algorithm (ACHA) cloud-top height, and Derived Motion Winds) that resolve structure outside the camera's field of view~\cite{Heidinger2020,Bresky2012,Daniels2019}. The $10$\,min ABI spine is far coarser than the DREAM cadence. A crossing of the scored field takes about $10$--$25$\,min, so ABI samples that crossing once or twice. The encoder consumes one hour of history, and the forecast leads stop at $30$\,min. At $15$--$20$\,m\,s$^{-1}$ a $30$\,min lead advects a parcel $27$--$36$\,km, which replaces the $10$--$25$\,km scored footprint with cloud that was outside it~\cite{Stuik2024,Velden2005}. Furthermore, GOES is not a transmission product. GOES only sees clouds in the thermal infrared. That view can provide rich correlative data, but it can be turned into optical attenuation only together with DREAM's measurements along summit sightlines~\cite{Stuik2024,Schmit2017}. On its own, GOES cannot map easily onto optical transmission.

\subsection{Problem Statement}
\label{sect:problem-intro}
Given the advancements already made, the objective is clear. We ask whether a predictive model can operate across multiple long time horizons and cover blind spots in the survey strategy. We forecast summit transmission using real DREAM and GOES data at leads that the scheduler consumes: 90\,s for the in-flight FBS queue, 5\,min for short-horizon dynamic pointing, 20\,min for a lower bound on pair-mode lookahead, and 30\,min as the longest trained lead (approximately the $\sim$33\,min WFD pair gap). The architecture of Sect.~\ref{sect:architecture} encodes the hour once and decodes each of those leads with shared weights.

\section{Methods}
\label{sect:methods}
We outline the development of a forecasting algorithm for transmission prediction. First, we describe the DREAM dataset in Sect.~\ref{sect:dream-data} and the satellite modality in Sect.~\ref{sect:goes-data}. Data preprocessing and the night-wise train, validation, and test split are in Sect.~\ref{sect:prep}. We then detail the classical baselines in Sect.~\ref{sect:baselines} and the deep-learning baselines in Sect.~\ref{sect:dl-baselines}. We then describe the proposed architecture in Sect.~\ref{sect:architecture}, including how it joins a 24-frame camera hour to slower unique-scan GOES and how a shared decoder emits several leads. We also detail how those networks, and the spherical ConvLSTM and SimVP baselines, are fitted. We then define the loss, including the yield mean-squared-error objective, the night-capped loader, the plateau schedule that trains each run to saturation, and the calibration warm start, in Sect.~\ref{sect:training}. We then describe the metrics used to gauge performance in Sect.~\ref{sect:metrics}. Finally, we experiment with data-modality ablations of the proposed model, one withholding GOES and one withholding DREAM, in Sect.~\ref{sect:ablations-methods}.

\subsection{DREAM Transmission Maps}
\label{sect:dream-data}
DREAM photometers catalog stars ($V<8.4$) every 6.4\,s and inverse-variance combine the five cameras into nested HEALPix extinction maps every $\sim$32\,s~\cite{Stuik2022,Stuik2024,Talens2018,Gorski2005}. We use the operational level-5 ($\mathrm{NSIDE}=32$) systematic-baseline product (\texttt{sys}), not the zeropoint product (\texttt{zps}). The \texttt{zps} maps tie each exposure to an instrumental zeropoint derived from that frame, so any zeropoint drift or miscalibration is absorbed into $A$ and looks like atmosphere. The \texttt{sys} maps instead reference a long-term MASCARA-style atmospheric baseline; they do not depend on the per-exposure zeropoint, and are therefore the cleaner estimate of gray attenuation per tile~\cite{Talens2018,Segev2026}. Only tiles above $\sim$20$^\circ$ altitude with $n_{\mathrm{obs}}\ge5$ are retained; the scores in Sect.~\ref{sect:metrics} use the cap above $30^\circ$. Transmission is
\begin{equation}
\label{eq:transmission}
T = 10^{-0.4\,A}.
\end{equation}
Raw $T$ still has tile-scale speckle. Photon counts in a HEALPix cell are low, the stellar sample is sparse, and scintillation adds fluctuations that are uncorrelated from tile to tile. A hard transmission cut on that field therefore traces noise islands rather than the decks a scheduler can steer around. Clouds that matter for a Rubin pointing span many tiles, so a modest spatial average suppresses the uncorrelated component while leaving those decks. We Gaussian-smear $T$ on the nested HEALPix faces with $\sigma=3.5$ face pixels ($\sim$6$^\circ$ at $\mathrm{NSIDE}=32$) before labeling, and we smooth the last-frame $T$ the same way, so the residual is not an artifact of smoothing only the future observed DREAM transmission. The same reasoning motivates few-cell Gaussian kernels on DREAM alt-az charts before motion correlation ($\sigma=5$ cells) and before cloud-boundary extraction ($\sigma=2.5$ cells)~\cite{Segev2026}. Fig.~\ref{fig:window-inputs}(a) is one hour of this smeared field, ending at $t_{\mathrm{end}}$.

We stage a tile as cloudy at $T\le0.63$ on the smoothed field. In magnitudes that is $A=0.5$, a $\sim$38\% reduction in transmitted flux ($10^{-0.4\times0.5}\approx0.63$). At that attenuation photometry is no longer photometric and a WFD visit through the tile is a wasted shutter. That is the cloudy definition we train and score against~\cite{Segev2026}.

For our work, the native DREAM cadence would produce on the order of $\sim 120$ frames per 1\,h sample. Using that native cadence made training and prototyping challenging with our compute, so we elected to use a linearly sampled subset of $24$ frames instead. This gives sufficient resolution to track slow-moving clouds and enough cadence contrast to demonstrate the model's multi-cadence strategy. 

Furthermore, we applied a Gaussian smoothing kernel to the DREAM transmission maps. This acts as a noise filter so that the algorithm sees coherent clouds rather than pixel speckles. Each extinction grid is smoothed with $\sigma = 5$ grid units, a standard Gaussian kernel similar to that in~\cite{Segev2026}. This choice was originally motivated in~\cite{Segev2026} by empirical tests. Fig.~\ref{fig:window-inputs}(a) shows a sample of 8 sequences of DREAM transmission maps to show how the Gaussian smoothing filter affects the maps and how the clouds evolve with time.

Lastly, we chose a DREAM daily archive of 305 complete Coordinated Universal Time (UTC) nights from 23 June 2025 to 28 May 2026. Windows from one night share that night's weather, so the split is by night rather than by time within a night, and every window from a night stays on one side. We sorted the nights by UTC date and indexed them from zero. Even-index nights are the development pool (train and validation). Odd-index nights are the test holdout. The index alternates. Consecutive archive nights are therefore placed on opposite sides of the cut. The split is not a run of contiguous multi-night blocks, and no buffer nights are dropped at the boundary. Of the 304 successive pairs in the sorted archive, 293 are one calendar day apart, and 141 of the 152 odd-index nights have both adjacent calendar dates in the even-index pool. Nights without a complete 1\,h history and $+20$\,min target are skipped. To create the validation nights from the even-index pool we assign $20\%$ of nights, evenly spaced through the year, to validation and the rest to training. The unused odd-index nights are the test holdout.

\subsection{Satellite Data}
\label{sect:goes-data}
DREAM measures attenuation close to the ground. We supplement its blind spots with an overhead satellite imager and a set of derived products. Since operations are at night, we use the night-capable GOES-East Advanced Baseline Imager (ABI) Full Disk products to see clouds in the infrared rather than the optical~\cite{Schmit2017,Stuik2024}. This data cube contains thermal imaging, water vapour, and high-altitude cloud geometry. 

We elected to use a number of these publicly available products. Thermal imaging produces C13 ($10.35\,\mu\mathrm{m}$), a measurement of cold-cloud brightness temperature and the $10$\,min time spine. C07 ($3.9\,\mu\mathrm{m}$) supplies night contrast for warm low water cloud that C13 does not see. C15 ($12.3\,\mu\mathrm{m}$), together with C13, is a split-window ice and optical-depth proxy. The differences C13$-$C07 and C13$-$C15 are therefore the night fog/stratus and thin-cirrus signals. There are also water-vapor channels C08, C09, and C10 ($6.2$, $6.9$, $7.3\,\mu\mathrm{m}$) which are upper-, mid-, and lower-tropospheric moisture upwind of the mountain~\cite{Schmit2017}.

On top of infrared imaging, we retrieve the ABI cloud mask (ACM). We also use the ABI Cloud Height Algorithm (ACHA), which records cloud-top height. This is accompanied by the ABI Cloud Top Phase (ACTP), a phase map used to split liquid and ice so that warm stratus and cirrus are not treated as the same class~\cite{Heidinger2009,Heidinger2012,Heidinger2020}. We also use the land surface temperature (LST) and total precipitable water (TPW), which supply slower column and surface context. 

\begin{figure}[htbp]
\centering
\includegraphics[width=\linewidth]{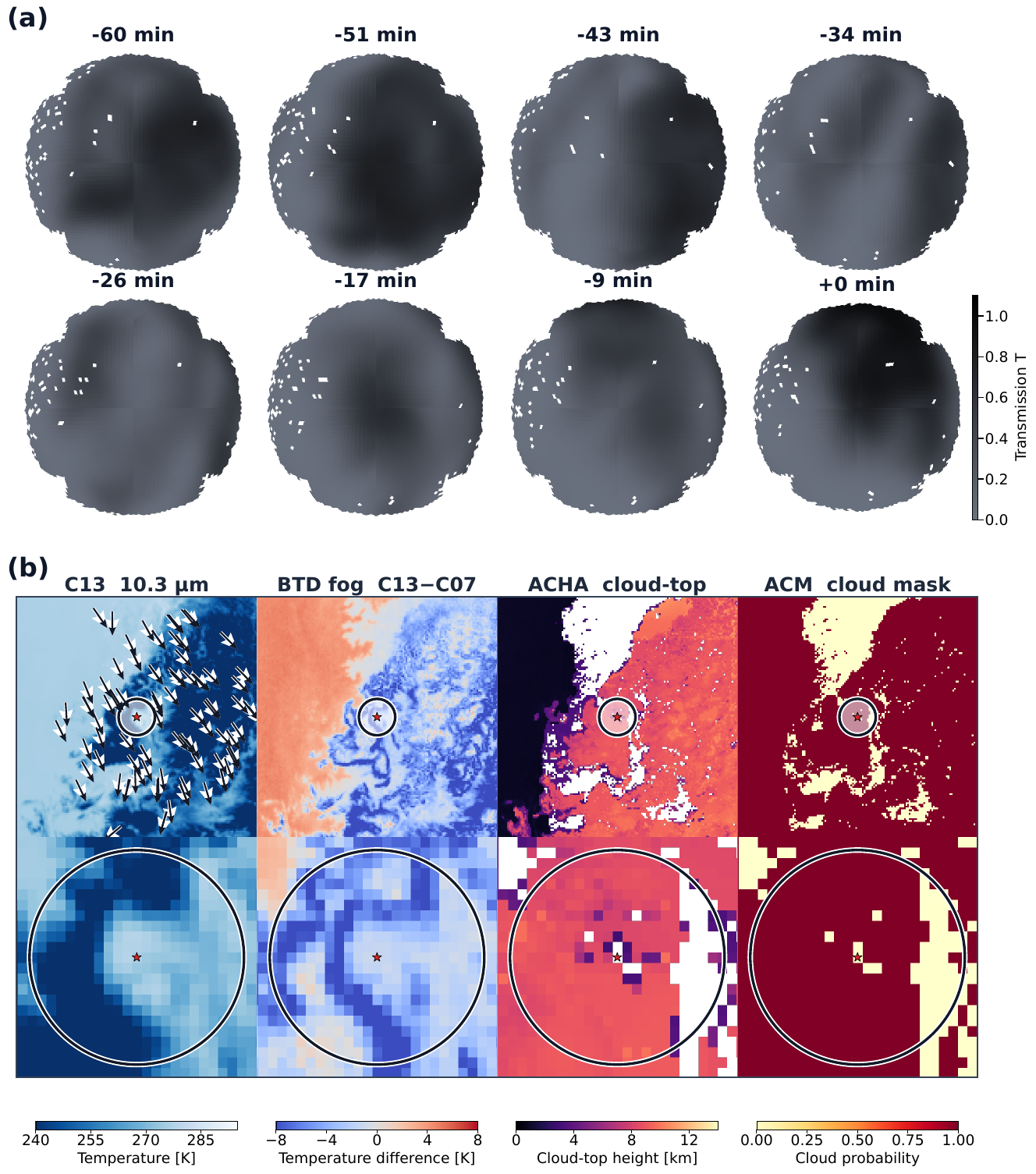}
\caption{One input window from the test holdout (21~May~2026).
(a)~Gaussian-smeared ($\sigma=3.5$) DREAM transmission over the hour ending at $t_{\mathrm{end}}$, shown as horizon-locked disks (north up) at the labeled offset from $t_{\mathrm{end}}$. The network consumes 24 frames at $\sim$2.5\,min; selected times are plotted. Dark red is $T\simeq 1$; black is opaque.
(b)~Coincident GOES-East ABI fields at $t_{\mathrm{end}}$: C13 ($10.3\,\mu\mathrm{m}$) brightness temperature, night fog/stratus difference C13$-$C07, ACHA cloud-top height, and the ACM cloud mask. The top row is the $\pm 165\,\mathrm{km}$ cutout around Cerro Pach\'on; the grey circle is a $25$\,km radius ground disk around the site (filled on the wide row), larger than the scored cap above $30^\circ$. Arrows are unit-direction Derived Motion Winds. The bottom row zooms to $1.12\times$ that disk.}
\label{fig:window-inputs}
\end{figure}

Lastly, we use the Derived Motion Winds (DMW), which are a layer-mean $(u,v)$ from nested correlation through an ABI triplet, with heights from ACHA. These are independent measurements rather than a camera-inferred velocity~\cite{Bresky2012,Daniels2019,Velden2005}. At typical high-cloud speeds of $\sim$15--20\,m\,s$^{-1}$, a $20$\,min step moves a parcel $\sim$18--24\,km. That distance is one crossing of the scored sky above $30^\circ$, about $10$--$25$\,km across for decks $1$--$7$\,km above the summit (Appendix~\ref{sect:app-buildup}), so a cloud at the zenith has left that cap in about $5$--$15$\,min. A high deck can still overlap the wider camera down to $\sim$20$^\circ$. Fig.~\ref{fig:window-inputs}(b) shows four of these products on one window---C13, C13$-$C07, ACHA, and ACM---with DMW direction arrows on the $\pm 165\,\mathrm{km}$ cutout and a zoom to a $25$\,km radius ground disk around the site.

\subsection{Data Preprocessing and Splitting}
\label{sect:prep}

Firstly, we need to preprocess DREAM transmission data. In each transmission map, tiles with $n_{\mathrm{obs}}<5$ or non-finite $T$ are masked and filled with 0. Transmission is clipped to $[0,3]$ and scaled on the training split to $z=(T-m)/q$, where $m$ is the median and $q$ the interquartile range. Only the last frame and the label receive the Gaussian smear of Sect.~\ref{sect:dream-data}. Furthermore, because DREAM maps are equatorial nested HEALPix, a fixed tile index is a sky position that rotates with local sidereal time. We reindex each frame onto a horizon sphere (zenith at the pole, north at $\phi=0$) so tile $k$ is a fixed sightline. Lastly, DREAM maps stellar photometry, so a bright moon above the horizon adds scattered light that looks like extinction. We therefore drop any window whose last frame has lunar illumination $\ge 0.25$ and geometric altitude $>0^\circ$, which aligns with~\cite{Segev2026}. The cut is per window, instead of per night, which means a night can still contribute its moon-dark hours; nights that are moonlit throughout are removed entirely. We also drop windows that are more than $99\%$ cloudy on observed tiles, which are uninformative for a nowcast. After these cuts the development set has 99 unique training nights ($52\,376$ windows) and 26 unique validation nights ($13\,322$ windows). The test holdout has 117 unique nights ($62\,524$ windows).

Secondly, the GOES satellite data are a $\pm1.5^\circ$ ($\sim$165\,km) Full Disk cutout around Cerro Pach\'on, kept at a native resolution of $\sim$2\,km ($135\times149$). In time, these images are aligned onto the C13 $10$\,min spine. Because the cadence is lower, a one-hour DREAM window holds only a handful of distinct ABI files, each repeated across several $\sim$2.5\,min camera slots. We form the filter differences C13$-$C07 and C13$-$C15 before filling missing infrared pixels, so a $270\,\mathrm{K}$ fill on one channel cannot create a false fog or split-window signal. GOES-16 views the summit at $35.6^\circ$ from
zenith, so a cloud at height $h$ is plotted several kilometres toward the satellite. We use ACHA heights to undo that shift before sampling the satellite map along each DREAM sightline. Each modality then receives its own normalization: ACM is clipped to $[0,1]$, ACTP is scaled by $[0,4]/4$, liquid or ice is a $0,1$ mask, LST is normalized as $(K-275)/15$, TPW as $(mm-2)/4$, $\Delta$C13 is clipped to $(\Delta K/15, 2)$, and $\Delta$ACM is clipped to $[-1,1]$. The complete data product is summarized in Tab.~\ref{tab:sat-cube}.

\begin{table}[t]
\centering
\caption{Satellite cube seen by the model (\texttt{sat}, 18 channels).
Infrared differences are formed in kelvin \emph{before} the $270\,\mathrm{K}$
fill. Missing C13/C07/C15/C08--C10 become $270\,\mathrm{K}$; missing ACHA
is $0\,\mathrm{m}$; LST/TPW use that frame's spatial median.}
\label{tab:sat-cube}
\small
\begin{tabular}{@{}lll@{}}
\toprule
Channel & Name & Normalisation \\
\midrule
\texttt{c13} & 10.3\,$\mu$m clean-window temperature & $(T_{10.3}-278)/15$ \\
\texttt{c07} & 3.9\,$\mu$m night water/stratus & $(T_{3.9}-278)/15$ \\
\texttt{c15} & 12.3\,$\mu$m dirty-window temperature & $(T_{12.3}-278)/15$ \\
\texttt{btd\_fog} & Night fog / stratus contrast & $(T_{10.3}-T_{3.9})/5$ \\
\texttt{btd\_split} & Thin-ice / split-window contrast & $(T_{10.3}-T_{12.3})/5$ \\
\texttt{acm} & Satellite cloud mask & ABI mask, clipped to $[0,1]$ \\
\texttt{acha} & Cloud-top height & $(h-62\,\mathrm{m})/1639\,\mathrm{m}$ \\
\texttt{phase} & Cloud-top phase class & $\mathrm{ACTP}/4$ (0 clear \ldots\ 4 ice) \\
\texttt{liquid} & Warm water-cloud flag & $1$ if $0.5\le\mathrm{ACTP}<2.5$ \\
\texttt{ice} & Ice / cirrus flag & $1$ if $\mathrm{ACTP}\ge 3.5$ \\
\texttt{lst} & Land skin temperature & $(T_\mathrm{skin}-272)/15$ \\
\texttt{tpw} & Column water vapour & $(w-2\,\mathrm{mm})/4\,\mathrm{mm}$ \\
\texttt{dc13} & 10-min clean-window change & $\Delta T_{10.3}/15$, clipped to $[-2,2]$ \\
\texttt{dacm} & 10-min mask change & $\Delta\mathrm{ACM}$, clipped to $[-1,1]$ \\
\texttt{c08} & 6.2\,$\mu$m upper water vapour & $(T_{6.2}-278)/15$ \\
\texttt{c09} & 6.9\,$\mu$m mid-level water vapour & $(T_{6.9}-278)/15$ \\
\texttt{c10} & 7.3\,$\mu$m low/mid water vapour & $(T_{7.3}-278)/15$ \\
\texttt{is\_missing} & Pre-fill infrared hole mask & $1$ where any of those infrared planes was NaN \\
\bottomrule
\end{tabular}
\end{table}

Together, these steps assemble the data product for training, validation, and test. Fig.~\ref{fig:window-inputs} is one truncated example: (a)~the camera hour and (b)~the GOES cutout at $t_{\mathrm{end}}$. 

\subsection{Classical Baselines}
\label{sect:baselines}
The classical baselines are: Eulerian persistence, optical flow, frozen-flow cross-correlation (the current dynamic scheduler), and DMW wind advection.

\paragraph{Eulerian persistence.}
Persistence is simple: we copy the last observed map, $\hat{T}(\hat{n},t_{\mathrm{end}}+H)=T(\hat{n},t_{\mathrm{end}})$~\cite{Germann2002,Yang2019,Chow2011}.

\paragraph{Dense optical flow.}
For the optical flow, the Lucas--Kanade model assumes brightness constancy on a locally translational image~\cite{Lucas1981,Bouguet2000}. Performing this on a sphere is challenging. The approach we took is to project the sphere onto a flat disk, measure the motion, slide the disk, and paint it back onto the sphere.  We take the last two transmission maps each sampled onto a $256\times256$ azimuthal-equidistant chart, zenith at the center, out to $80^\circ$ zenith distance; this makes the disk the camera footprint. We then perform a two-level pyramidal Lucas--Kanade fit (window of 15 chart pixels), which yields a dense $(u,v)$ in chart pixels per pair. The latest chart is semi-Lagrangian advected by $(H/\Delta t_{\mathrm{obs}})(u,v)$: each target pixel looks back along the flow and bilinearly samples the last chart~\cite{Staniforth1991,Bowler2006,Pulkkinen2019}. The advected disk is then bilinearly resampled onto nested HEALPix at the target local sidereal time. Pixels whose backtrace left the disk, and solver failures, are filled from the last equatorial map.

\paragraph{Frozen-flow cross-correlation.}
Next we implemented a version of Segev's production tracker, because the original does not estimate a dense field. It finds one global translation of a coarse alt-az chart and uses that vector to pick the next Rubin pointing~\cite{Segev2026}. We keep that estimator and change only the output. The same computed velocity advects the full transmission map to $H=20$\,min so it can be scored against the other full-sky nowcasts.

The published scheduler works on DREAM's cartesian alt-az product. We start from equatorial HEALPix, so consecutive maps are first sampled onto a $30\times30$ zenithal chart that matches that grid. After a Gaussian smooth of $\sigma=5$ cells, every integer shift of $\pm10$ cells is scored with a Pearson coefficient on the overlapping valid pixels (minimum overlap 50 cells). Shifts with correlation below 0.5 are dropped. Surviving vectors from the last five frames of the window are converted to chart pixels per second and averaged. That single $(v_x,v_y)$ is then applied as a spatially constant flow: the latest chart is semi-Lagrangian advected over $H$ and remapped to nested HEALPix at the target local sidereal time, with the same off-disk and failure fallback to persistence as optical flow. 

\paragraph{Derived Motion Winds (DMW) advection.}
DMW advection is a classical method incorporating another data modality. Here DMW are an independent meteorological $(u,v)$, not a velocity fitted from DREAM~\cite{Bresky2012,Daniels2019,Velden2005}. They are Earth-relative, so advection is a ground-frame backtrace rather than a chart-pixel warp. For each night we pool quality-controlled ABI wind vectors within $200$\,km of Cerro Pach\'on and above $4$\,km geometric height, take the layer-mean $(u,v)$ and median height $h$ per $10$\,min scan, and use the latest scan at or before $t_{\mathrm{end}}$. A missing file, an empty pool, or a vanishing vector reverts to persistence.

The last transmission map is then displaced by $(\Delta x,\Delta y)=(u,v)\,H$. For every target tile at $t_{\mathrm{end}}+H$ we convert its sightline to a ground position (east, north) where it pierces a plane-parallel layer at $h$, step that position back by $(\Delta x,\Delta y)$, and read $T$ from the HEALPix tile that observed the source location at $t_{\mathrm{end}}$. Sightlines below $10^\circ$ altitude are left undefined, because the ground range diverges. Inflow from outside the camera footprint has no source tile and is filled from the last map. The $20$\,min step at typical high-cloud speeds of $\sim$15--20\,m\,s$^{-1}$ moves a parcel $\sim$18--24\,km, one crossing of the scored sky above $30^\circ$, so much of that cap is inflow the camera has not seen. The method cannot invent cloud that has not yet entered the camera~\cite{Germann2002,Chow2011,Stuik2024}.

\subsection{Deep Learning Baselines}
\label{sect:dl-baselines}
We first outline the Chebyshev layer, a common mechanism in the baseline models and the proposed model, and one that remains novel for nowcasting relative to the original literature. We then describe the ConvLSTM and SimVP models, which use the Chebyshev layer and act as literature baselines.

\paragraph{Chebyshev layers} were developed to force representations to obey the spherical geometry of the input data. This makes them useful for learning from transmission maps, which live natively on the sphere, as opposed to rectilinear images, which suit standard convolutions. There is no rectilinear transformation of features for HEALPix data; instead, tiles are represented by a \textit{graph}. Each nested HEALPix tile is a vertex. The HEALPix data structure provides its neighbors, up to eight of them. The weight of each edge on the graph is a Gaussian of the chord distance between pixel centers. From those weights we construct the weighted adjacency matrix $A$. We care about how neighbors differ from each other, so we compute a normalized Laplacian $L = \mathbb{1} - D^{-1/2}A D^{-1/2}$, where $D$ is the degree matrix (the sum of the edge weights at each tile). The Laplacian records how a value at a point differs from its neighbors. The layer aggregates neighboring tiles, one ring at a time. The Chebyshev order sets the reach of that ring of aggregation: the farthest ring is $K-1$ hops (Eq.~\ref{eq:cheb}). The features $X$ are mixed by $T_k(\tilde{L})$, and $\theta_k$ is a learned weight for that ring.

\begin{equation}
  y
  = \sum_{k=0}^{K-1} \theta_k \, T_k(\tilde{L})\, X,
  \qquad
  \tilde{L} = L - \mathbb{I} = -D^{-1/2} A D^{-1/2}.
  \label{eq:cheb}
\end{equation}
The rescaling puts the eigenvalues of $\tilde{L}$ in $[-1,1]$. Chebyshev polynomials are a stable basis on that interval, applied by the recurrence in Eq.~\ref{eq:poly}.
\begin{equation}
  T_0(X) = X,
  \qquad
  T_1(X) = \tilde{L} X,
  \qquad
  T_k(X) = 2\tilde{L}\, T_{k-1}(X) - T_{k-2}(X).
  \label{eq:poly}
\end{equation}

\paragraph{ConvLSTM} implements a standard ``peephole cell'' of Shi et~al. It is unchanged except that every convolution $W\ast X$ is a Chebyshev convolution (ChebConv)~\cite{Shi2015}. Hidden features and cell states live on the sphere, with shape $(B,N_{\mathrm{pix}},H)$ and $H=57$. We then stack two cells and unroll them over the history; a $K=1$ ChebConv reads the last hidden state as $\Delta\tau_0$. Width is set so the recurrent baseline has $1.19\times10^5$ parameters, approximately similar in size to our proposed model. Recurrence is the implicit advection memory: the cell can move, grow, or fade structure that a frozen-flow tracker can only translate.

\paragraph{SimVP} drops recurrence. SimVP takes the 24 frames, stacked on the channel axis ($24\times2=48$ channels) and mixed by ChebConv~\cite{Gao2022}. A two-level encoder maps $\mathrm{NSIDE}=32\to16\to8$ with hidden width 32, then 64 at the bottleneck. Four residual ChebConv blocks translate at $\mathrm{NSIDE}=8$ ($768$ tiles). The decoder upsamples with skip additions from the encoder. A $K=1$ head emits $\Delta\tau_0$ at full resolution. This is the feed-forward baseline we train for the numbers in Sect.~\ref{sect:benchmark}: $1.17\times10^5$ parameters, approximately matched in capacity to the proposed model, camera history only. ConvLSTM is the recurrent counterpart under the same spherical layers and residual persist head. Both are trained from scratch on the recipe of Sect.~\ref{sect:training}. Each baseline head emits $\Delta\tau_0$ and is decoded by the same Beer--Lambert update as the proposed model (Eq.~\ref{eq:beer-lambert}), with the per-tile gate and the residual scale held at $1$.

\subsection{Model Architecture}
\label{sect:architecture}
The architecture of our model comes with a variety of design choices to optimally leverage our multimodal dataset. We first designed custom encoders for the spherical DREAM camera data, the satellite imagery, the vector point-cloud wind data, and the numerical metadata. After the encoder, we employ geometry fusion to combine features from flat geometry with spherical geometry and unite the modalities. We also developed methods of stitching the data onto the same temporal backbone by mixing features to handle multi-cadence input. 
\begin{figure}[ht]
    \centering
    \includegraphics[width=\textwidth]{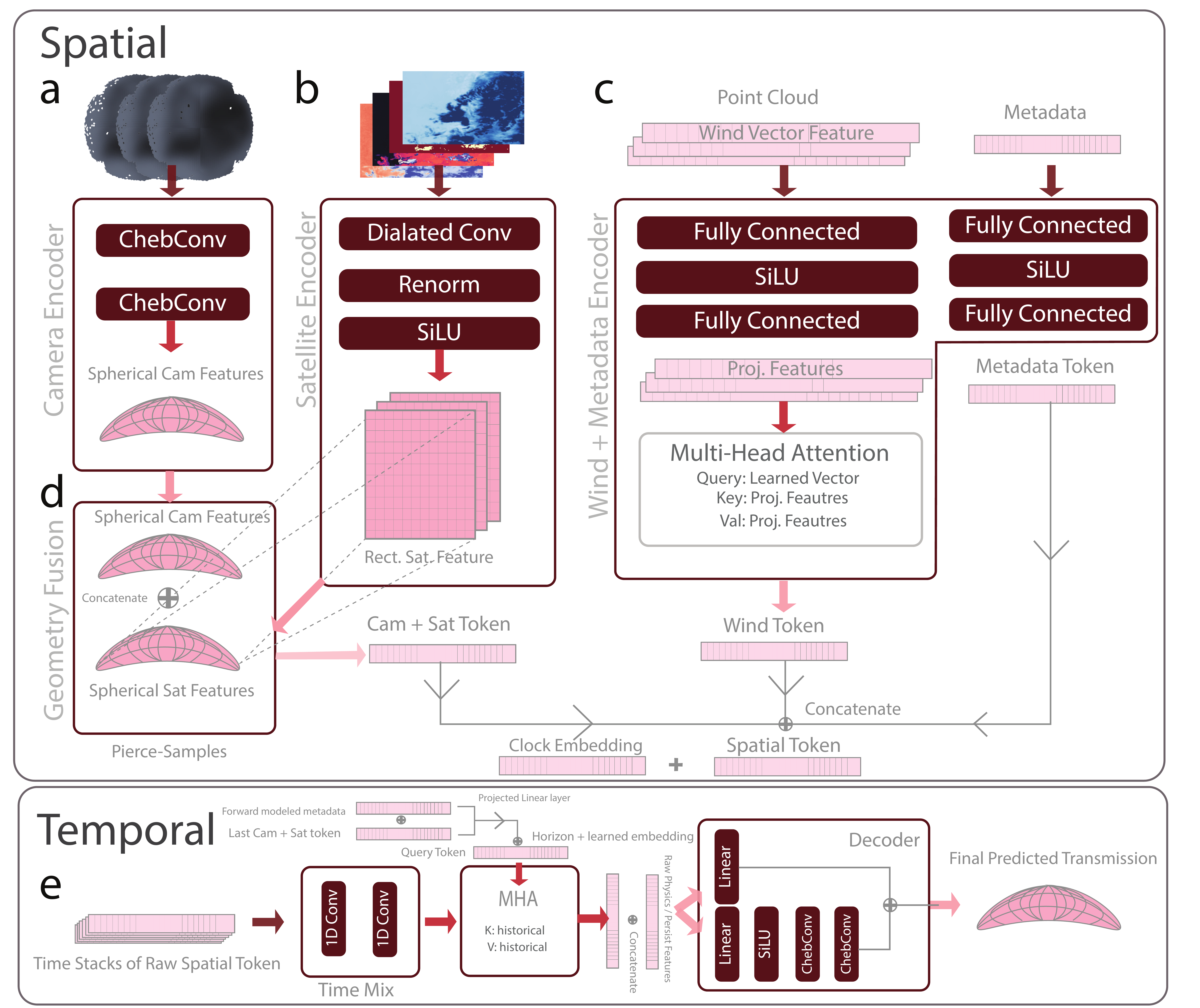}
    \caption{Architecture of the proposed multi-modal transmission forecasting model.
    The spatial block fuses four modalities:
    (a)~a camera encoder of 24 horizon-locked sky maps by residual ChebConv;
    (b)~a satellite encoder of native GOES ABI patches by dilated 2-D convolution, run once per unique $10$\,min ABI scan and gathered onto the 24 camera times;
    (c)~a wind and metadata encoder, with multi-head attention over DMW point clouds and an MLP on hour and scan age;
    (d)~geometry fusion (pierce join) that samples the rectangular satellite features onto spherical sightlines at cloud-height decks.
    The temporal block (e) mixes the 24-frame history with a forecast query by a masked dilated 1-D convolution and a lead-conditioned attention pool. History is encoded once. A shared decoder then emits a residual on last-frame optical depth at each requested lead (90\,s, 5\,min, 20\,min, 30\,min), using that lead's embedding and wind-displaced GOES pierce. The output updates a persistence baseline via the Beer--Lambert law (Eq.~\ref{eq:beer-lambert}).}
    \label{fig:transmission_arch}
\end{figure}

\paragraph{Camera encoder.}
The first component is the camera encoder for DREAM transmission. We ingest individual frames and use the camera encoder as a spatial embedder. Each frame produces 7 features per node (tile) in the graph convolution. The first feature is $T$, then the first and second differences $\Delta T$ and $\Delta^{2}T$ (zero where either frame is missing, so a hole is not motion), and $(\sin a,\cos a,\sin\psi,\cos\psi)$ with $a$ the tile altitude and $\psi$ the azimuth from north. A valid-tile flag is concatenated as an eighth channel. A two-block DeepSphere ChebConv CNN of hidden width 16 ($K=3$; Sect.~\ref{sect:dl-baselines}) embeds each of the 24 frames independently, producing a 16-dimensional vector at every tile.

\paragraph{Satellite encoder.}
\label{sect:multicadence}
Similar to the camera encoder, we use a separate satellite encoder, because the camera and satellite clocks and geometries disagree. DREAM frames span the hour at $\sim$2.5\,min, while C13 Full Disk updates every $10$\,min, so consecutive camera slots often carry identical ABI data. To address this, we pack each sample's distinct scans and encode that satellite stack \textit{once} with a rectilinear stride-1 dilated 2-D CNN (kernels $3,5,3$; dilations $1,2,4$; widths $32\to64\to64$; group normalization and SiLU after each convolution). We then gather the geometry-fused features back onto the 24 camera times. 

\paragraph{Winds and metadata.}
Unlike satellite and DREAM data, wind data introduces yet another new data modality. DMW are a sparse point cloud, rather than a simple grid: up to 400 quality-controlled vectors per frame, each $(u,v,h,x,y,\mathrm{age})$. Because the number of wind vectors varies per frame and they lack a fixed spatial structure, a multi-head attention (MHA) block is used to aggregate them. However, in the MHA the query vector is learned rather than tied to the dataset, because the wind data are unordered, unlike a usual sequence, and so have no natural query token to attend to. The learned query asks the set for the flow that matters, and the attention picks those points. This query vector attends over the valid points (four-head scaled-dot-product attention) and is concatenated with a masked mean and max pool, then linearly projected to a 24-dimensional token per frame. This design allows the model to summarize an arbitrary and unordered set of wind vectors into a coherent, fixed-size scene-level wind token.

The metadata are tabular (local solar hour, layer-mean winds, scan age, and a scalar lead placeholder; nine scalars) and are passed through a two-layer MLP to 16 dimensions. We strip key identifiers from the metadata, holding day of year at zero; otherwise the even/odd night split would let the network memorize seasonal weather from the calendar. Both metadata and point-cloud features are later broadcast to every tile. The decoder query is a second copy of that wind set and metadata, forward-modeled from fields known at $t_{\mathrm{end}}$. The point cloud is the last history frame, with $(u,v,h,x,y,\mathrm{age})$ left unchanged. Layer-mean winds in the query metadata are those same issue-time values. Local solar hour is the hour at the camera frame originally matched as the $15$\,min target (within $25\%$ of $15$\,min), and scan age is the age of the latest C13 scan on the last history frame plus that same lag. Day of year remains zero. The lead placeholder on this shared copy is $\tanh(H/6\,\mathrm{h})$ at $H=20$\,min. One copy is shared across the four decoded leads. The lead requested at decode time enters through the embedding of Eq.~\ref{eq:lead-delta} and through the upwind GOES pierce, which reads the image at $t_{\mathrm{end}}$ after a displacement $(u,v)H$. 

\paragraph{Geometry fusion (Pierce join).}
We currently treat each geometry natively; however, the features must eventually merge. Thus we have to perform a fusion of geometric representations. Fundamentally the camera tiles live on the sphere while ABI pixels live on a tangent plane. To fuse these together, we trace for each sightline along the sphere and compute the east/north ground position where the ray meets plane-parallel decks at $2.5$, $6$, and $10$\,km, the same intersection used by DMW advection in Sect.~\ref{sect:baselines}. These deck heights are sampled because the height at which a ray actually hits is not known ahead of time. Sightlines below $10^\circ$ altitude are left undefined (the ground range diverges). The satellite CNN feature map is then bilinearly sampled at those three heights. The mix is applied to the CNN features at every history time, and to five raw channels at $t_{\mathrm{end}}$: C13$-$C07, C13$-$C15, C10, ACM, and liquid-phase fraction. This gives the model a sense of what the current state at that time looks like on a sphere.

To assist the model, we also guess what will sit on a camera sightline a time $H$ ahead, and we feed that into the network. We do not only read the GOES pixel overhead; we also step upwind by the distance the wind moves in that time. The wind is the average GOES motion from the last two scans (only points that actually have a wind hit). That upwind pixel is where today's air (and any cloud still off the camera's view) is sitting right now. We read two things there: the last sat-CNN features, and the five raw GOES channels. At training time, we randomly hide satellite and wind (5\% each) and metadata (20\%) so the model can still run on the camera if GOES is missing.

\paragraph{Temporal mixer.}
To assemble individual time-step features into a cohesive, time-dependent data product, we temporally mix the features. To address this, we first need to mix the asynchronous clock of different input modalities as discussed previously. We do so by copying data from the slowest input features to make up for the much faster DREAM cadence. Once all the spatial data has been encoded, at each history time the 16-D camera embed and the 64-D pierced CNN features are linearly fused to 64 dimensions, then concatenated with the broadcast wind and metadata tokens and projected back to 64. A Fourier timestamp embedding of frame age and $\Delta t$ is added, so the mixer sees clock time rather than frame index. The result is one token per tile per frame. 

After constructing the tokens we pass them into a masked dilated 1-D convolution along the 24-frame axis, independently at every tile (kernel 3, dilations $1,2,4,8$, GELU). This mixes neighbouring features in time, to account for the sparse updates of slower-clocked inputs such as the satellite images. 

This temporally mixed sequence of tokens is then passed into a multi-head attention block. The keys and values are the historical tokens, except for the most recent token, and the query is the most recent token plus the horizon embedding. This produces a context feature, which is added back to the most recent token through a residual gated by a single-layer MLP $g$. 
\begin{equation}
\label{eq:hist-skip}
\mathrm{skip}=\mathrm{last}+\tanh\bigl(g(\mathrm{last})\bigr)\cdot\mathrm{context}.
\end{equation}
This design provides skip connections, handles the asynchronous satellite cadence, and lets the model select context according to the requested horizon. 

\paragraph{Multi-horizon decode.}
\label{sect:multihorizon}
To achieve multi-horizon prediction within a single monolithic model, we chose one-shot decoding rather than an autoregressive method. This is because errors can easily pile up on the way to 20--40\,min. An autoregressive rollout would also require generating the entire satellite and wind cube, not only the transmission map, which is expensive both in offline training and in real-time inference. It is also difficult to generate data that remain valid and do not poison the late-time rollout. 

We designed a single-shot head which takes a flag for the four horizons, 90\,s, 5\,min, 20\,min, and 30\,min. This flag enters the internal kinematic model and a small two-layer MLP embedder. The embedder takes $\bigl(\log(H/H_{\mathrm{ref}}),\,H/H_{\mathrm{ref}}\bigr)$ with $H_{\mathrm{ref}}=20$\,min. The last layer is zero-initialized, and the vector added at decode is the centered difference
\begin{equation}
\label{eq:lead-delta}
e(H)=\mathrm{MLP}\bigl(\phi(H)\bigr)-\mathrm{MLP}\bigl(\phi(20\,\mathrm{min})\bigr).
\end{equation}
We nominally set the default task to a 20\,min lookahead, so that lead is the easy query by construction. 

This token $e(H)$ is used in two places. First, it is added to the most recent spatial token and used in the query attention. Second, it is added to the last token in the skip connection. 

\paragraph{Decoder and Beer--Lambert residual.}
The forecast has to return a transmission map, and the geometry of that map is part of the decoder. A deck that thickens by the same amount everywhere is a single change in vertical cloud. From the camera, that change is a different transmission on every tile, because a lower altitude looks through a longer path. The decoder therefore emits the change in natural zenith optical depth, and the Beer--Lambert law converts that change back to transmission,
\begin{equation}
\label{eq:beer-lambert}
T = \mathrm{e}^{-\tau/\sin a},\qquad \tau = \tau_0 + \alpha\, g\, \Delta\tau.
\end{equation}
Here $\tau_0$ is the zenith optical depth inverted from the last DREAM frame, $\Delta\tau$ is the residual from the decoder, $\alpha$ is a learned scale, and $a$ is the altitude of the tile. The factor $g$ is a per-tile gate, so a quiet tile can stay on the measured sky. The gate is
\begin{equation}
\label{eq:gate}
g=\sigma\bigl(w^{\top}[p_i^{\mathrm{now}}\| p_i^{\mathrm{up}}]+b\bigr).
\end{equation}
The vector $p_i^{\mathrm{now}}$ is the raw satellite sample from the latest scan, read where that spherical sightline meets the flat GOES image. The vector $p_i^{\mathrm{up}}$ is that same image read upwind, at the point the measured wind carries onto the sightline over the lead. Those two vectors pass through a learned linear layer, and a sigmoid maps the result to $g$.

\subsection{Training}
\label{sect:training}
To keep the comparison uniform, the baselines, the proposed model, and the ablations share one training setup, so that all runs are fairly graded.

\paragraph{Loss function}
Our primary loss function, shared by all models, is an airmass-weighted mean squared error (MSE) on transmission $T$, which we define as $\mathcal{L}_{yield}$. 
\begin{equation}
  \mathcal{L}_{\mathrm{yield}}(\hat T, T)
  = \frac{\sum_{i} \sin a_i \, (\hat T_i - T_i)^2}
         {\sum_{i} \sin a_i}.
\end{equation}
We average the losses across horizons with equal weight. We also define a false-clear loss $\mathcal{L}_{FC}$.

\begin{equation}
  \mathcal{L}_{\mathrm{FC}}
  = \frac{1}{|C|}
    \sum_{i \in C}
    \mathrm{ReLU}(\hat T_i - \tau),
  \qquad
  C = \{\, i : T_i \le \tau,\; \text{valid},\; a_i \ge 30^\circ \,\}.
\end{equation}
\begin{equation}
  \mathcal{L}_{\mathrm{risk}}
  = \mathcal{L} + \lambda \, \mathcal{L}_{\mathrm{FC}}.
\end{equation}
This false-clear term is not active for the whole of training. It comes into effect only after the yield loss has saturated. That is, $\lambda$ depends on the epoch and remains zero for most of the run. Its value is then set to $\lambda = 0.01$.

\paragraph{Training Curriculum}
We discovered that for optimal training performance we should not be training with the false-clear loss active from the start, because it can block convergence of the reconstruction loss. The model sometimes buys a better false-clear score by biasing the sky toward cloud, which is the safer hedge. We therefore train in two phases. The first uses only $\mathcal{L}_{yield}$. When learning saturates (the validation loss plateaus four times), we finetune for $25$ epochs at a lower learning rate with $\mathcal{L}_{FC}$ turned on. 

\subsection{Metrics}
\label{sect:metrics}
We use several metrics to gauge the model against the baseline methods.

\paragraph{Mean absolute yield error (MAYE).}
We know that the photon yield at fixed exposure is $N_{\mathrm{phot}}\propto T$ (Eq.~\ref{eq:transmission}), so a squared error on transmission is a squared error on collected photons. The train loss is the airmass-weighted mean squared error of $\hat T$ against the observed DREAM transmission at $t_{\mathrm{end}}+H$, the square of the MAYE residual in Sect.~\ref{sect:metrics}:
\begin{equation}
\label{eq:yield-mse}
\mathcal{L}
=\frac{\sum_i w_i\bigl(\hat T_i-T_i\bigr)^2}{\sum_i w_i},
\qquad
w_i=
\begin{cases}
\sin a_i & a_i\ge 30^\circ \text{ and tile $i$ observed},\\
0 & \text{otherwise}.
\end{cases}
\end{equation}

\paragraph{False-clear penalty.}
One of the most critical metrics is the false-clear rate. It records when we suggest an observable patch while the sky is in fact cloudy. In effect, it measures time wasted. We set a tile as cloudy at $T\le0.63$ ($A=0.5$; Sect.~\ref{sect:dream-data}). The false-clear rate is the fraction of those cloudy tiles that the forecast called clear,
\begin{equation}
\label{eq:false-clear}
\eta_{\mathrm{FC}}=\frac{\mathrm{FN}}{\mathrm{TP}+\mathrm{FN}},
\end{equation}
where $\mathrm{FN}$ counts cloudy tiles the forecast called clear and $\mathrm{TP}$ counts cloudy tiles it called cloudy, on tiles observed at both $t_{\mathrm{end}}$ and $t_{\mathrm{end}}+H$ with $a\ge30^\circ$. This is $1-\mathrm{recall}$, a miss the scheduler cannot recover. In pair-mode that wastes the first visit of the pair as well as the second~\cite{Jones2018,Schwamb2023}. 

\paragraph{Five-sigma survey depth error $\Delta m_{5\sigma}$.}
The five-sigma point-source limiting magnitude is the depth metric used internally by the Rubin scheduling community~\cite{Ivezic2019,Segev2026}. Under the signal-to-noise ratio (SNR) proxy $\mathrm{SNR}=\sqrt{N_{\mathrm{phot}}}$,
\begin{equation}
\label{eq:m5}
m_{5\sigma}=20-2.5\log_{10}\bigl(5/\sqrt{N_{\mathrm{phot}}}\bigr).
\end{equation}
The zeropoint $20$ is a convenient scale; it cancels in a difference. Because $N_{\mathrm{phot}}\propto T$, the depth implied by the prediction and by the true map differ by
\begin{equation}
\label{eq:dm5}
\Delta m_{5\sigma}
=m_{5\sigma}(\hat T)-m_{5\sigma}(T)
=1.25\log_{10}(\hat T/T).
\end{equation}
Before this ratio, and before $D_{\mathrm{eff}}$ in Eq.~\ref{eq:deff}, both $\hat T$ and the observed DREAM transmission are clipped to $[0,1]$, and each argument of the logarithm is floored at $10^{-4}$. The decoder already confines $\hat T$ to $[0.05,0.999]$, so the metric floor moves only that observed future map. A positive $\Delta m_{5\sigma}$ is an overestimate of depth: the scheduler expected more photons than arrived. The full LSST $m_5$ also depends on seeing, sky brightness, and airmass~\cite{Ivezic2019}; holding those fixed, extinction enters only through $N_{\mathrm{phot}}$. Coadded depth over the ten-year survey scales the same way with accumulated photons, $\Delta m_{5\sigma}^{\mathrm{coadd}}=1.25\log_{10}f_{\mathrm{eff}}$, so a persistent per-visit error compounds over thousands of visits~\cite{Segev2026}.

\paragraph{Effective telescope diameter $D_{\mathrm{eff}}$.}
Atmospheric loss is therefore equivalent to shrinking the aperture. If $f$ is the factor by which cloud reduces the photon budget relative to a photometric visit ($f=1/T$), Rubin's $8.4$\,m primary~\cite{Ivezic2019} becomes
\begin{equation}
\label{eq:deff}
D_{\mathrm{eff}}=D\sqrt{1/f}=8.4\,\mathrm{m}\times\sqrt{T}.
\end{equation}
Segev notes that pointing through avoidable cloud on a patchy night shrinks that mirror to $\sim$6.4--6.5\,m relative to an omniscient clear-patch pointing~\cite{Segev2026}. We apply the same translation to a forecast. The predicted map implies $D_{\mathrm{eff}}(\hat T)$ and the true map $D_{\mathrm{eff}}(T)$, so the residual is meters of glass. Comparing the photon yield on tiles the model called clear against the yield on the true clear mask converts a mask mistake into those units.

\subsection{Ablations}
\label{sect:ablations-methods}
We perform data ablations to test whether multimodality is what allows the forecast to reach each horizon. We ablate data modalities rather than architecture, because the gain hinges on fusing modalities rather than on the architectural choice. 

\paragraph{No satellite.}
We drop the satellite modality by setting the GOES-East ABI cube of Sect.~\ref{sect:goes-data} and the Derived Motion Winds to zero. The 24 DREAM frames and the last-frame persist residual $\tau_0$ remain. This effectively makes the proposed architecture run as a camera-only nowcast. 

\paragraph{No DREAM input.}
We also drop the DREAM sequence by setting the 24 stored transmission frames, their first and second differences, and the valid-tile mask to zero. Altitude and azimuth on the camera encoder remain. In the decoder skip, last-frame transmission is the constant $1$ and the last difference is $0$. GOES features in the same skip vector are unchanged, so the head has no measured camera map to copy. This tests whether ABI is a substitute for on-site photometry or only an upwind context for it.

\section{Results}
\label{sect:results}
We outline the results on the held-out nights. Firstly, we compare the proposed model with the classical and neural baselines at each trained lead in Sect.~\ref{sect:benchmark}: the $90$\,s queue, then the $5$, $20$, and $30$\,min horizons, and where that lead appears as cloud builds, clears, or stays put. We then test the two modality drops of Sect.~\ref{sect:ablations-methods}, withholding GOES and withholding DREAM, in Sect.~\ref{sect:ablations-results}.

\subsection{Benchmark Model Performance}
\label{sect:benchmark}
We report scores on the subset of the odd-night holdout whose window-mean observed DREAM transmission falls in $T\in[0.1,0.9]$: $11{,}691$ moon-dark windows. This is the operationally interesting band and aligns with~\cite{Segev2026}. Windows above $0.9$ are photometric and every method trivially agrees, while windows below $0.1$ are completely clouded in and no pointing is available. The band retains the patchy and thin-cloud conditions that a dynamic scheduler could exploit. All learned entries are four independent seeds trained from scratch. The four classical baselines are deterministic given the cache, so their seed spread is exactly zero. A $\pm$ value or a shaded band on a learned model is the standard deviation across those four seeds. Whiskers on a median are the interquartile range of the windows in that bin. Tab.~\ref{tab:horizons_maye_fc} tracks MAYE and $\eta_{\mathrm{FC}}$ at each trained lead. Tab.~\ref{tab:mean_t_01_09} is the $H=20$\,min and $H=30$\,min scorecard with the remaining depth and aperture metrics.
\begin{table}[ht]
\caption{MAYE and false-clear rate $\eta_{\mathrm{FC}}$ versus lead on the $T\in[0.1,0.9]$ holdout. Mean $\pm$ 1 standard deviation over four seeds; classical methods are deterministic. A window is dropped at a lead if the observed DREAM frame at that lead is missing, so $n$ is $11{,}680$ (90\,s), $11{,}671$ (5\,min), $11{,}691$ (20\,min), and $11{,}530$ (30\,min). Lower is better. Best value in each column is bold.}
\label{tab:horizons_maye_fc}
\centering
\resizebox{\textwidth}{!}{%
\begin{tabular}{lcccccccc}
\hline
 & \multicolumn{2}{c}{90\,s} & \multicolumn{2}{c}{5\,min} & \multicolumn{2}{c}{20\,min} & \multicolumn{2}{c}{30\,min} \\
Model & MAYE & $\eta_{\mathrm{FC}}$ & MAYE & $\eta_{\mathrm{FC}}$ & MAYE & $\eta_{\mathrm{FC}}$ & MAYE & $\eta_{\mathrm{FC}}$ \\
\hline
Eulerian persistence & 0.0455 & 0.065 & 0.0845 & 0.122 & 0.1587 & 0.235 & 0.1777 & 0.269 \\
Dense optical flow & 0.0459 & 0.064 & 0.0855 & 0.118 & 0.1598 & 0.233 & 0.1795 & 0.264 \\
Frozen flow (cross-corr) & \textbf{0.0390} & 0.047 & 0.0786 & 0.100 & 0.1601 & 0.224 & 0.1810 & 0.251 \\
Lagrangian wind advection & 0.0464 & 0.064 & 0.0868 & 0.119 & 0.1622 & 0.226 & 0.1824 & 0.253 \\
ConvLSTM & 0.0766 $\pm$ 0.0044 & 0.151 $\pm$ 0.035 & 0.1000 $\pm$ 0.0028 & 0.194 $\pm$ 0.026 & 0.1557 $\pm$ 0.0021 & 0.298 $\pm$ 0.029 & 0.1742 $\pm$ 0.0014 & 0.319 $\pm$ 0.031 \\
SimVP & 0.0496 $\pm$ 0.0013 & 0.053 $\pm$ 0.019 & 0.0836 $\pm$ 0.0041 & 0.119 $\pm$ 0.031 & 0.1737 $\pm$ 0.0152 & 0.425 $\pm$ 0.150 & 0.1987 $\pm$ 0.0139 & 0.516 $\pm$ 0.154 \\
Proposed & 0.0479 $\pm$ 0.0008 & \textbf{0.045} $\pm$ 0.008 & \textbf{0.0766} $\pm$ 0.0007 & \textbf{0.083} $\pm$ 0.015 & \textbf{0.1267} $\pm$ 0.0013 & \textbf{0.174} $\pm$ 0.031 & \textbf{0.1502} $\pm$ 0.0025 & \textbf{0.191} $\pm$ 0.037 \\
\hline
\end{tabular}
}
\end{table}

At short timescales, classical cross-correlation wins at 90 seconds. As expected from the current FBS dynamic scheduler, immediate lookups by cross-correlation improve on persistence, cutting MAYE by $14\%$ and $\eta_{\mathrm{FC}}$ by $28\%$ (Tab.~\ref{tab:horizons_maye_fc}). Nearly all other models, except ConvLSTM ($68\%$ worse on MAYE), track persistence. The proposed model is better on the false-clear rate ($31\%$ below persistence), although that seed spread reaches the winning frozen-flow value.

\begin{figure}[htbp]
\centering
\includegraphics[width=\linewidth]{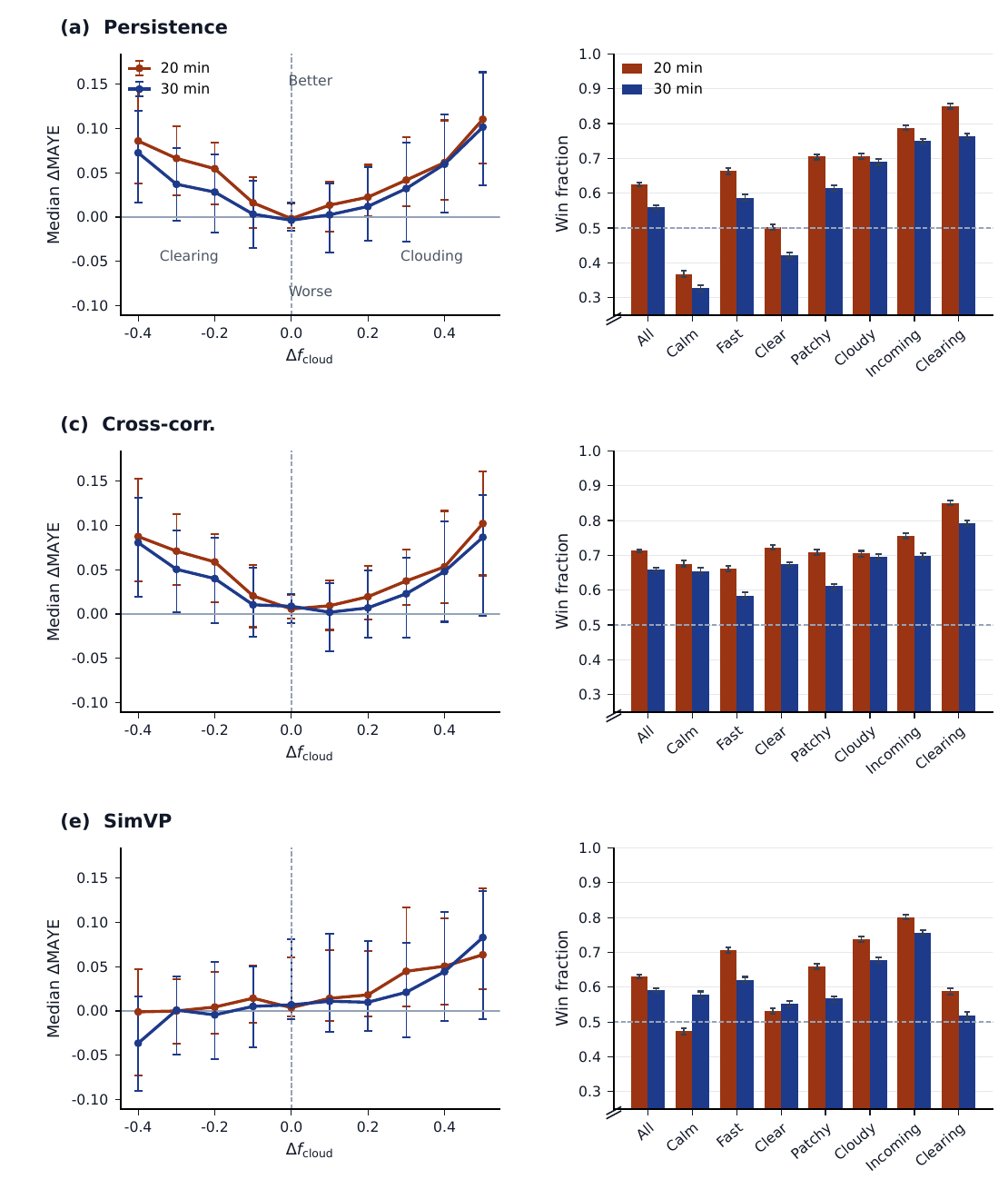}
\caption{Per-window skill of the proposed model versus Eulerian persistence (a), the deployed frozen-flow cross-correlation tracker (c), and SimVP (e) on the same holdout as Tab.~\ref{tab:mean_t_01_09} (mean observed DREAM transmission $T\in[0.1,0.9]$). Left: median $\Delta\mathrm{MAYE}=$ baseline $-$ Proposed versus the change in cloudy fraction $\Delta f_{\mathrm{cloud}}$ (positive is incoming cloud). Whiskers are the interquartile range of windows in the bin. Right: fraction of windows where Proposed has lower MAYE than that baseline, with 16--84\% bootstrap intervals on that fraction. The eight bars are defined in the text; a window can fall in more than one. Brown is $H=20$\,min; navy is $H=30$\,min. The win-fraction axis is broken below $\sim$0.3.}
\label{fig:skill-regimes}
\end{figure}

At longer timescales, the proposed model dominates. After 90\,s, longer-horizon prediction becomes a challenge for the classical models and for the other nowcasting networks. The proposed model then leads in both MAYE and the false-clear rate ($9\%$ / $32\%$ below persistence at $5$\,min, $20\%$ / $26\%$ at $20$\,min, and $15\%$ / $29\%$ at $30$\,min; Tab.~\ref{tab:horizons_maye_fc}). 
\begin{table}[ht]
\caption{Pooled performance on holdout windows with mean transmission $T\in[0.1,0.9]$. Top: $H=20$\,min ($11{,}691$ windows). Bottom: $H=30$\,min ($11{,}530$ windows). Mean $\pm$ 1 standard deviation over four training seeds; the four classical baselines are deterministic. $\Delta m_{5\sigma}$ is the signed depth bias and $|\Delta m_{5\sigma}|$ the per-tile depth error, both in magnitudes. MAE is the mean absolute error of $D_{\mathrm{eff}}$. Lower is better for every column except the signed bias, where zero is best. Best value in each block and column is bold.}
\label{tab:mean_t_01_09}
\centering
\resizebox{\textwidth}{!}{%
\begin{tabular}{lcccccc}
\hline
\multicolumn{7}{c}{$H=20$\,min} \\
\hline
Model & MAYE & yield MSE & $\eta_{\mathrm{FC}}$ & $\Delta m_{5\sigma}$ & $|\Delta m_{5\sigma}|$ & $D_{\mathrm{eff}}$ MAE (m) \\
\hline
Eulerian persistence & 0.1587 & 0.0534 & 0.235 & \textbf{$-$0.0150} & 0.2316 & 1.024 \\
Dense optical flow & 0.1598 & 0.0537 & 0.233 & $-$0.0207 & 0.2341 & 1.032 \\
Frozen flow (cross-corr) & 0.1601 & 0.0533 & 0.224 & $-$0.0273 & 0.2357 & 1.036 \\
Lagrangian wind advection & 0.1622 & 0.0543 & 0.226 & $-$0.0304 & 0.2387 & 1.049 \\
ConvLSTM & 0.1557 $\pm$ 0.0021 & 0.0488 $\pm$ 0.0016 & 0.298 $\pm$ 0.029 & $+$0.0948 $\pm$ 0.0111 & 0.2062 $\pm$ 0.0017 & 0.974 $\pm$ 0.011 \\
SimVP & 0.1737 $\pm$ 0.0152 & 0.0590 $\pm$ 0.0112 & 0.425 $\pm$ 0.150 & $+$0.1487 $\pm$ 0.0328 & 0.2221 $\pm$ 0.0139 & 1.071 $\pm$ 0.082 \\
Proposed & \textbf{0.1267} $\pm$ 0.0013 & \textbf{0.0329} $\pm$ 0.0013 & \textbf{0.174} $\pm$ 0.031 & $+$0.0497 $\pm$ 0.0180 & \textbf{0.1690} $\pm$ 0.0033 & \textbf{0.791} $\pm$ 0.011 \\
\hline
\multicolumn{7}{c}{$H=30$\,min} \\
\hline
Model & MAYE & yield MSE & $\eta_{\mathrm{FC}}$ & $\Delta m_{5\sigma}$ & $|\Delta m_{5\sigma}|$ & $D_{\mathrm{eff}}$ MAE (m) \\
\hline
Eulerian persistence & 0.1777 & 0.0669 & 0.269 & \textbf{$-$0.0005} & 0.2588 & 1.145 \\
Dense optical flow & 0.1795 & 0.0673 & 0.264 & $-$0.0054 & 0.2609 & 1.156 \\
Frozen flow (cross-corr) & 0.1810 & 0.0673 & 0.251 & $-$0.0196 & 0.2660 & 1.169 \\
Lagrangian wind advection & 0.1824 & 0.0682 & 0.253 & $-$0.0216 & 0.2683 & 1.179 \\
ConvLSTM & 0.1742 $\pm$ 0.0014 & 0.0600 $\pm$ 0.0018 & 0.319 $\pm$ 0.031 & $+$0.1030 $\pm$ 0.0129 & 0.2346 $\pm$ 0.0011 & 1.094 $\pm$ 0.007 \\
SimVP & 0.1987 $\pm$ 0.0139 & 0.0762 $\pm$ 0.0128 & 0.516 $\pm$ 0.154 & $+$0.1752 $\pm$ 0.0292 & 0.2547 $\pm$ 0.0110 & 1.222 $\pm$ 0.070 \\
Proposed & \textbf{0.1502} $\pm$ 0.0025 & \textbf{0.0451} $\pm$ 0.0023 & \textbf{0.191} $\pm$ 0.037 & $+$0.0491 $\pm$ 0.0261 & \textbf{0.2009} $\pm$ 0.0047 & \textbf{0.935} $\pm$ 0.018 \\
\hline
\end{tabular}
}
\end{table}
We then evaluate the medium and long horizons, particularly 20--30\,min, where paired visits need protection (Tab.~\ref{tab:mean_t_01_09}). The proposed model improves on the classical models, the deployed tracker, and the other neural models. At 20\,min, MAYE / $\eta_{\mathrm{FC}}$ are $20\%$ / $26\%$ below persistence, $21\%$ / $22\%$ below the deployed frozen-flow tracker, $19\%$ / $42\%$ below ConvLSTM, and $27\%$ / $59\%$ below SimVP. The same improvement appears in the yield MSE and in the $D_{\mathrm{eff}}$ MAE: $38\%$ / $23\%$ below persistence and $33\%$ / $19\%$ below ConvLSTM. At $30$\,min the MAYE / $\eta_{\mathrm{FC}}$ cuts are $15\%$ / $29\%$ over persistence, $17\%$ / $24\%$ over frozen flow, and $14\%$ / $40\%$ over ConvLSTM. Fig.~\ref{fig:skill-regimes} shows those MAYE gains against incoming cloud and by wind and cloud regime.

The eight bars on the right of Fig.~\ref{fig:skill-regimes} are slices of those same windows, where a window is counted in every bar it meets. "All" is every scored window. "Calm" and "Fast" are the bottom and top quartiles of the median Derived Motion Winds speed in the window. "Clear", "Patchy", and "Cloudy" are the last-frame mean transmission: $T(t_0)\ge 0.75$, $0.40\le T(t_0)<0.75$, and $T(t_0)<0.40$. Incoming is a drop of at least $0.08$ in that mean, $T(t_0)-T(t_0+H)\ge 0.08$, and Clearing is a rise of at least $0.08$.

\begin{figure}[htbp]
\centering
\includegraphics[width=\linewidth]{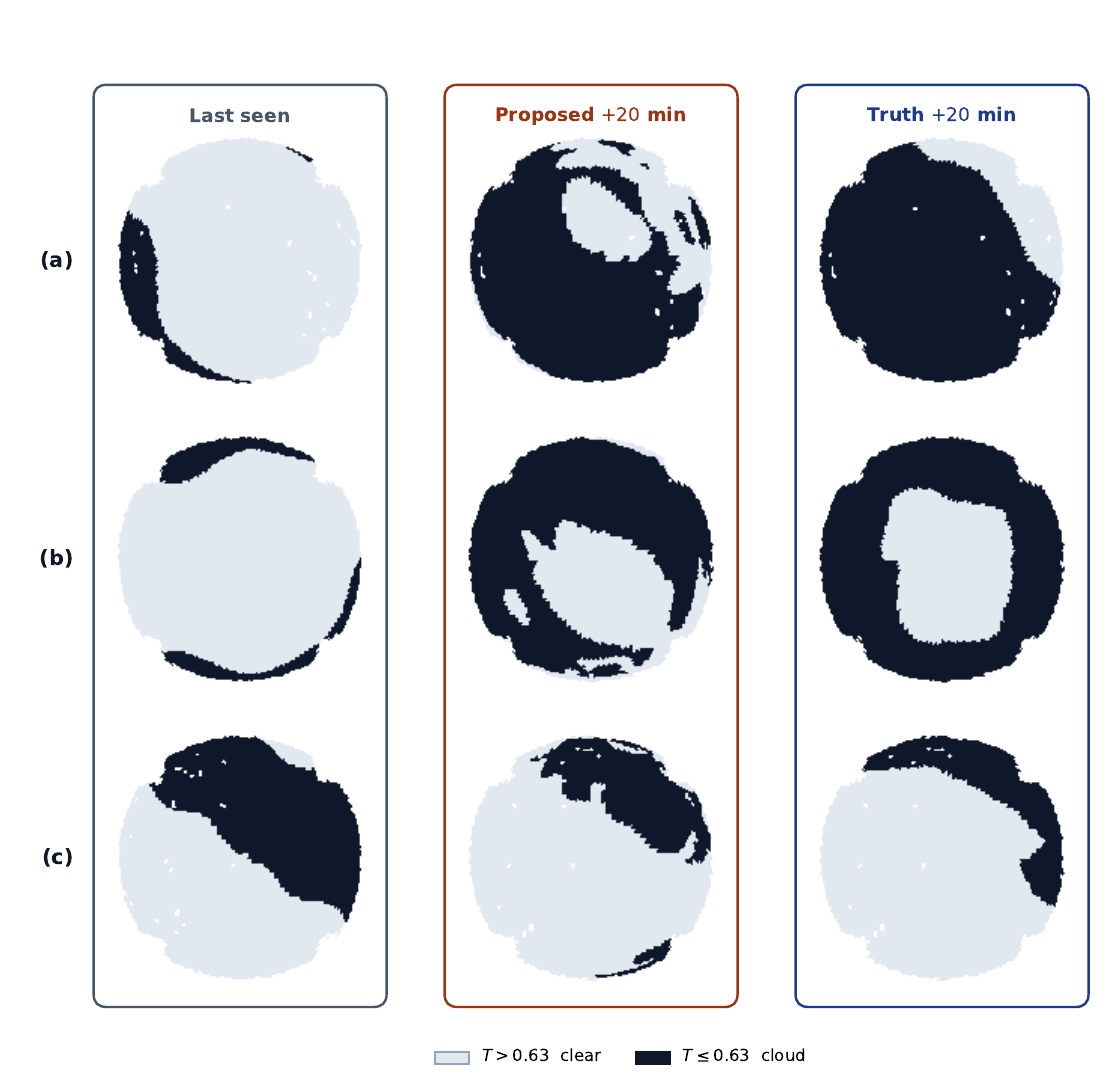}
\caption{Example $+20$\,min cloud masks ($T\le 0.63$) on the same holdout as Tab.~\ref{tab:mean_t_01_09}. Horizon-locked disks, north up. Boxed columns are last seen (last DREAM frame at $t_0$), Proposed, and camera truth. (a),(b)~incoming cloud on two nights. (c)~clearing.}
\label{fig:forecast-successes}
\end{figure}

The proposed model performs best, relative to the baselines, when the weather changes most (Fig.~\ref{fig:skill-regimes}(a),(c)). At $20$\,min, on incoming-cloud windows ($T(t_0)-T(t_0+H)\ge 0.08$) the proposed model beats persistence and cross-correlation on $79\%$ / $76\%$ of windows. On clearing windows ($T(t_0)-T(t_0+H)\le -0.08$) those win rates rise to $85\%$ / $85\%$. The comparison with SimVP is asymmetric (Fig.~\ref{fig:skill-regimes}(e)). The model's skill in clearing situations is statistically the same as SimVP (win rates $59\%$ / $52\%$ at $20$ / $30$\,min), while in incoming-cloud situations its performance is higher (win rates $80\%$ / $76\%$). We see that neural models can learn the dynamics already visible in the camera, as the clearing scores show. However, incoming cloud is limited by which modalities are available. Fig.~\ref{fig:forecast-successes} shows that split on the maps. In (a) and (b), incoming nights, the last DREAM frame is still mostly clear and twenty minutes later a deck has entered; the forecast follows the observed DREAM map at $t_0+H$ rather than copying the frame at $t_0$. Panel (c) is the clearing case where a north-east deck already in the camera slides off, and the same head tracks the retreat.
We also check how the model responds when the environment does not change. The proposed model meets the baseline in that case. When the change in cloud fraction reaches zero, the predictive advantage essentially disappears, and the model often fails to beat persistence, as Fig.~\ref{fig:skill-regimes}(a) shows for calm weather. On nearly static windows ($|\Delta f_{\mathrm{cloud}}|<0.04$) it wins on only $42\%$ / $38\%$ of windows at $20$ / $30$\,min, and on the calm-wind quartile the win rate falls to $37\%$ / $33\%$, versus $63\%$ / $57\%$ over the full holdout. 

\subsection{Ablations}
\label{sect:ablations-results}
\begin{figure}[htbp]
\centering
\includegraphics[width=\linewidth]{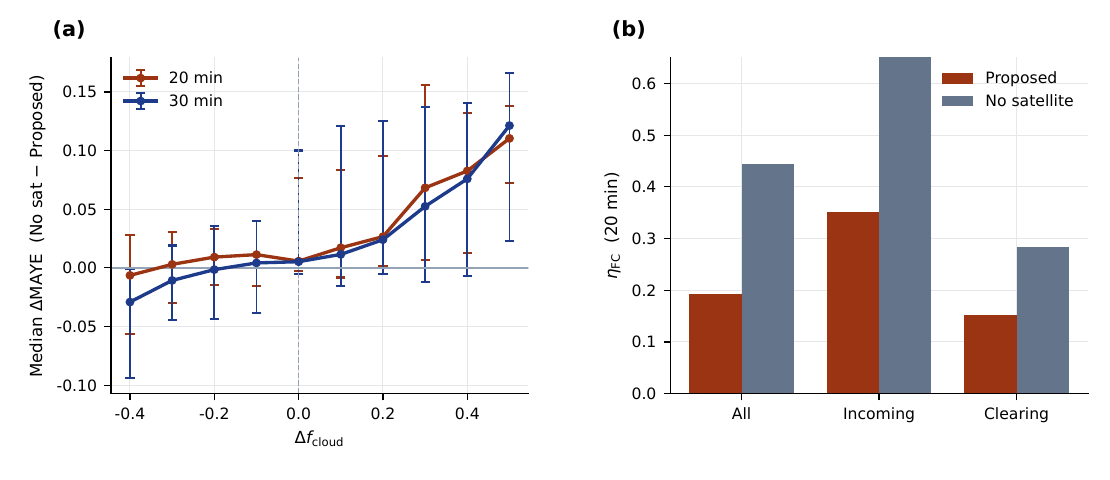}
\caption{Satellite ablation on the same holdout as Tab.~\ref{tab:mean_t_01_09}. (a) Median $\Delta\mathrm{MAYE}=$ No satellite $-$ Proposed versus the change in cloudy fraction $\Delta f_{\mathrm{cloud}}$ (positive is incoming cloud). Whiskers are the interquartile range of windows in the bin. Red is $H=20$\,min; navy is $H=30$\,min. (b) False-clear rate $\eta_{\mathrm{FC}}$ at $H=20$\,min on all windows, incoming windows ($\Delta f_{\mathrm{cloud}}\ge 0.08$), and clearing windows ($\Delta f_{\mathrm{cloud}}\le -0.08$).}
\label{fig:ablation-satellite}
\end{figure}

Our ablation results show that satellite imagery is critical. Tab.~\ref{tab:ablations} shows that dropping GOES satellite imagery greatly raises MAYE, and performance falls to levels similar to Lagrangian flow or frozen flow. At $20$\,min, No satellite MAYE is $32\%$ higher than Proposed and within $3$--$5\%$ of persist / frozen flow (Tab.~\ref{tab:mean_t_01_09}); $\eta_{\mathrm{FC}}$ is more than double Proposed and $68\%$ above persist. At $30$\,min MAYE is $27\%$ above Proposed and $6$--$8\%$ above persist / frozen flow, while $\eta_{\mathrm{FC}}$ is $160\%$ above Proposed. This is expected: at those horizons the model has no way to anticipate new cloud flowing in, and its predictive capacity is fundamentally limited. Fig.~\ref{fig:ablation-satellite} shows that the missing foresight is the incoming cloud. The median MAYE gap stays near zero on static and clearing windows and opens once $\Delta f_{\mathrm{cloud}}$ is positive ($\Delta f_{\mathrm{cloud}}\ge 0.08$). On those incoming windows at $20$\,min, $\eta_{\mathrm{FC}}$ more than doubles (Fig.~\ref{fig:ablation-satellite}(b)), while clearing stays much closer.

\begin{figure}[htbp]
\centering
\includegraphics[width=\linewidth]{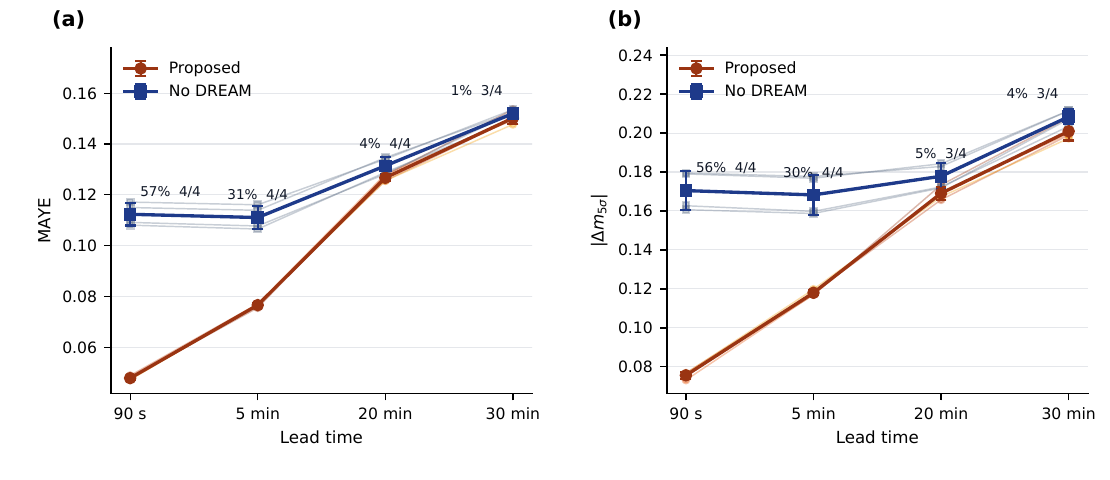}
\caption{DREAM ablation versus lead time on the same holdout as Tab.~\ref{tab:ablations}, four independently trained seeds. Faint traces are individual seeds; markers and bars are the mean $\pm$ one seed standard deviation. Lower is better. (a) MAYE. (b) Absolute limiting-depth error $|\Delta m_{5\sigma}|$. Labels give the mean reduction relative to No DREAM and the number of seeds in which Proposed has lower error.}
\label{fig:ablation-dream}
\end{figure}

At the same time, the ablations show that DREAM supplies the near-term state. Dropping the on-site camera data does not collapse long-horizon skill the way dropping GOES does, because the satellite still supplies upwind cloud. Fig.~\ref{fig:ablation-dream} shows where DREAM does matter: the first few minutes. Across four seeds, removing DREAM raises MAYE at both $90$\,s and $5$\,min, and all four seeds are worse in both cases. Relative to those camera-free scores, Proposed is lower by $57\%$ and $31\%$. The same comparison raises $|\Delta m_{5\sigma}|$ by $56\%$ and $30\%$. By the paired-visit leads in Tab.~\ref{tab:ablations} the gap has already closed to $3.6\%$ MAYE at $20$\,min (four of four seeds) and $1.2\%$ at $30$\,min (three of four). Absolute depth error follows the same decay ($4.9\%$ / $3.6\%$). DREAM is therefore the high-cadence photometric anchor of the current sky, while GOES increasingly determines skill once new cloud has time to enter the field.

Tab.~\ref{tab:ablations} repeats the $H=20$\,min and $H=30$\,min scorecard of Tab.~\ref{tab:mean_t_01_09} for the proposed model and the two modality drops of Sect.~\ref{sect:ablations-methods}.
\begin{table}[ht]
\caption{Modality ablations on the same holdout as Tab.~\ref{tab:mean_t_01_09}. Top: $H=20$\,min ($11{,}691$ windows). Bottom: $H=30$\,min ($11{,}530$ windows). Mean $\pm$ 1 standard deviation over four training seeds. No satellite zeros GOES ABI and Derived Motion Winds. No DREAM zeros the camera transmission channels and writes $T=1$ into the skip. Lower is better for every column except the signed bias, where zero is best. Best value in each block and column is bold.}
\label{tab:ablations}
\centering
\resizebox{\textwidth}{!}{%
\begin{tabular}{lcccccc}
\hline
\multicolumn{7}{c}{$H=20$\,min} \\
\hline
Model & MAYE & yield MSE & $\eta_{\mathrm{FC}}$ & $\Delta m_{5\sigma}$ & $|\Delta m_{5\sigma}|$ & $D_{\mathrm{eff}}$ MAE (m) \\
\hline
Proposed & \textbf{0.1267} $\pm$ 0.0013 & 0.0329 $\pm$ 0.0013 & 0.174 $\pm$ 0.031 & \textbf{$+$0.0497} $\pm$ 0.0180 & \textbf{0.1690} $\pm$ 0.0033 & \textbf{0.791} $\pm$ 0.011 \\
No satellite & 0.1670 $\pm$ 0.0036 & 0.0540 $\pm$ 0.0025 & 0.395 $\pm$ 0.039 & $+$0.1431 $\pm$ 0.0100 & 0.2166 $\pm$ 0.0038 & 1.037 $\pm$ 0.021 \\
No DREAM & 0.1314 $\pm$ 0.0034 & \textbf{0.0323} $\pm$ 0.0016 & \textbf{0.160} $\pm$ 0.020 & $+$0.0773 $\pm$ 0.0320 & 0.1778 $\pm$ 0.0066 & 0.828 $\pm$ 0.028 \\
\hline
\multicolumn{7}{c}{$H=30$\,min} \\
\hline
Model & MAYE & yield MSE & $\eta_{\mathrm{FC}}$ & $\Delta m_{5\sigma}$ & $|\Delta m_{5\sigma}|$ & $D_{\mathrm{eff}}$ MAE (m) \\
\hline
Proposed & \textbf{0.1502} $\pm$ 0.0025 & 0.0451 $\pm$ 0.0023 & 0.191 $\pm$ 0.037 & \textbf{$+$0.0491} $\pm$ 0.0261 & \textbf{0.2009} $\pm$ 0.0047 & \textbf{0.935} $\pm$ 0.018 \\
No satellite & 0.1913 $\pm$ 0.0039 & 0.0695 $\pm$ 0.0033 & 0.496 $\pm$ 0.041 & $+$0.1719 $\pm$ 0.0077 & 0.2489 $\pm$ 0.0037 & 1.185 $\pm$ 0.021 \\
No DREAM & 0.1520 $\pm$ 0.0017 & \textbf{0.0428} $\pm$ 0.0008 & \textbf{0.186} $\pm$ 0.026 & $+$0.0844 $\pm$ 0.0354 & 0.2084 $\pm$ 0.0039 & 0.959 $\pm$ 0.015 \\
\hline
\end{tabular}
}
\end{table}

\section{Discussion}
\label{sect:discussion}
Throughout this work we have demonstrated that neural networks can learn the dynamics of short-term weather patterns and can supersede handcrafted forward models at horizons greater than $\mathcal{O}(1\,\mathrm{min})$. Multimodality covers blind spots in both short-term and long-term prediction, which is why both satellite imagery and a high-cadence all-sky camera are needed. The experiments match that reading of the data and of the problem. 

\subsection{Impacts on Future Surveys}
\label{sect:survey-impact}
We give an order-of-magnitude estimate of the gain from a hypothetical deployment. About 80\% of survey time goes to WFD pairs. Within that share, historical DREAM data show that about 17\% of nights fall in the $T\in[0.1,0.9]$ range on which the models were benchmarked. On that subset the false-clear rate improves by $6\%$ in absolute terms. A naive product of those factors is a $\sim 1\%$ gain in survey capacity that would otherwise be lost to cloud. 

\subsection{Failure Modes and Caveats}
\label{sect:failure-impact}
Despite the demonstrated success in providing improved predictive power to current models, there are still failure modes and caveats in our design that need to be acknowledged. 

\paragraph{The 90\,s in-flight queue still prefers simple cross-correlation.} The same table shows the proposed model winning at $5$--$30$\,min, but losing at the horizon that sits in the immediate FBS queue. Frozen-flow cross-correlation has MAYE $0.0390$ against Proposed $0.0479\pm0.0008$; Proposed only matches it on $\eta_{\mathrm{FC}}$ (Tab.~\ref{tab:horizons_maye_fc}). ConvLSTM is $68\%$ worse than persistence on that lead, and SimVP is statistically consistent with persistence. We do not treat the shared decoder as a replacement for the deployed Segev tracker on the $\sim$90\,s in-flight batch.

\paragraph{Misses are not only a calm-weather problem.} In Fig.~\ref{fig:skill-regimes} the results already show that on nearly static windows ($|\Delta f_{\mathrm{cloud}}|<0.04$) median $\Delta\mathrm{MAYE}$ versus persist is $-0.002$ / $-0.004$ at $20$ / $30$\,min (win $42\%$ / $38\%$), and on the calm-wind quartile it is $-0.003$ / $-0.005$ (win $37\%$ / $33\%$). That is not the whole loss set. Over the full $T\in[0.1,0.9]$ holdout Proposed still has the worse MAYE on $37\%$ of $20$\,min windows and $43\%$ of $30$\,min windows. Even the incoming-cloud slice, where the median gain is largest ($+0.043$, win $79\%$ at $20$\,min), leaves about one in five windows worse than persist. The long-horizon headline is therefore an average of large clearing and incoming gains against a long tail of individual misses, including nights where cloud is arriving rather than sitting still.

\paragraph{DREAM can raise the pair-visit error.} Dropping the camera dataset still collapses the near term: Proposed MAYE is $57\%$ below the camera-free run at $90$\,s and $31\%$ below it at $5$\,min (Fig.~\ref{fig:ablation-dream}). At $20$ and $30$\,min the same ablation is no longer a uniform loss. No DREAM has the better yield MSE and the better $\eta_{\mathrm{FC}}$ in both blocks of Tab.~\ref{tab:ablations} ($0.0323$ / $0.160$ versus $0.0329$ / $0.174$ at $20$\,min; $0.0428$ / $0.186$ versus $0.0451$ / $0.191$ at $30$\,min). The full model keeps only a $3.6\%$ and $1.2\%$ MAYE edge. Once GOES has time to see the incoming deck, forcing the decoder through a one-hour camera sequence can slightly worsen the false-clear and yield scores. DREAM is the photometric anchor of the current sky; it is not automatically helpful at the pair-visit leads.

\paragraph{Alternating nights share multi-day weather.} The even/odd index leaves consecutive calendar nights on opposite sides of the development/test cut, so a marine-stratus deck or an upper-tropospheric moisture plume that lasts more than one night can enter both pools. After the lunar cut and the $>99\%$ cloudy-window cut, 117 odd-index nights remain in the holdout and 125 even-index nights remain in development (99 train, 26 validation). Of those 117 holdout nights, 93 still have both calendar-adjacent dates in that development set, 22 have one, and two (30 July 2025 and 31 January 2026) have none. Those two nights are too few to measure the leakage. Validation nights are taken from the even-index pool, so where the archive has no gap the nights next to them are odd-index holdout nights. A benchmark on contiguous multi-week blocks, with the nights at each block boundary removed, was not run, so the tables do not show scores on weather that the development nights never saw.

\subsection{Future Deployment Pathway and Online Training Plan}
\label{sect:deployment}
Deploying a data-driven model requires maintenance, as any other software or hardware system does. That maintenance has to be regular if the data distribution drifts, especially under a changing climate. Routine training or finetuning should therefore run during the day, before observations, on recent nights. A simple scheme finetunes the model and weights recent observations more heavily than the historical DREAM archive, so the prior tracks the latest weather. The training reported here weights nights equally. In operations, recent nights are likely to be more informative than older ones. 

\section{Conclusion}
\label{sect:conclusion}
We developed a single multimodal forecast of summit transmission that consumes a DREAM hour and GOES-East ABI and emits the four scheduler leads of 90\,s, 5\,min, 20\,min, and 30\,min. On the $T\in[0.1,0.9]$ holdout test set the proposed model is the best MAYE and false-clear entry from 5\,min onward. At the 20\,min pair-visit lead it cuts transmission error by $21\%$ and false-clear misses by $22\%$ versus the deployed frozen-flow tracker, and by $26\%$ versus persistence on false-clear. Those gains concentrate on nights when the sky is changing. Incoming cloud is the regime the camera alone cannot see, and dropping GOES returns the long-horizon scores to classical levels. Ablations split the two sensors the other way as well: DREAM is the photometric anchor of the first few minutes, while satellite imagery supplies the incoming-cloud signal once the lead is long enough for a new deck to enter the field.

The same tables also bound the claim. Frozen-flow still outperforms on the 90\,s in-flight queue on MAYE. On nearly static windows the model often loses to a copy of the last frame, and even the incoming-cloud slice leaves about one in five windows worse than persist. At 20--30\,min, dropping DREAM can slightly improve the false-clear and yield scores, so the camera is not automatically helpful at pair-visit leads. The holdout is the odd-index nights of an alternating split: 93 of the 117 scored test nights still have both calendar neighbors in the development pool, so a multi-day system can sit on both sides of the cut. A naive product of the 80\% WFD pair share, the 17\% of nights in the $0.1$--$0.9$ band, and the $6\%$ absolute false-clear improvement sits at $\sim 1\%$ of survey capacity. Any operational use would still need regular daytime finetuning on recent nights so the weights track the weather that is actually arriving.

\subsection*{Disclosures}

The authors declare that there are no financial interests, commercial affiliations, or other potential conflicts of interest that could have influenced the objectivity of this research or the writing of this paper.

\subsection*{Code, Data, and Materials Availability}

The data, the trained model, and the code will be made public when the paper is accepted for publication. An advance copy of the code is available on request to the corresponding author.

\subsection*{Acknowledgments}

P.X.M.\ and J.S.B.\ are supported by Schmidt Sciences (award G-25-70164). P.X.M.\ is also funded by the Canada Graduate Research Scholarship -- Doctoral program (CGRS D-609387-178) from the Natural Sciences and Engineering Research Council of Canada (NSERC). This work also used Google Cloud research credits for large-language-model API access. We thank Tiago Ribeiro for access to the DREAM data, and the DREAM team for preparing those datasets and for discussion, in particular Remko Stuik. We thank Lynne Jones for early discussions of the scheduler, and Chris Stubbs and Larom Segev for sharing advance copies of their work and thesis on the cross-correlation dynamic scheduler, which informed much of the early analysis. P.X.M.\ also thanks the LSST-DA Data Science Fellowship Program, which is funded by LSST-DA, the Brinson Foundation, the WoodNext Foundation, the Research Corporation for Science Advancement Foundation, and Kevin Wells; his participation in the program has benefited this work. The project began there as a hack-day exercise in data visualisation, which led to a closer look at the data and then to the realisation that it could address the core problem of forecasting transmission. P.X.M.\ also thanks the new friends made through the DSFP program and the many shenanigans that come from these sessions.

Large language models were used heavily in code development and in writing and proof-editing of the manuscript. Models include GPT5.6 Sol, DeepSeekV4, Claude Opus 5.5, GLM5.3, Kimi K3, Gemini 3.1 Pro, etc. The research directions, the ideas for executing the analysis, the evaluations, and the majority of the writing and interpretation of the results are the work of the authors.

\appendix
\section{Clear calls safer than the cloud cover}
\label{sect:app-clear-calls}

A further test compares $P(\text{true cloud} \mid \text{predicted clear})$ with the cloud fraction $f_{\mathrm{cloud}}$. The goal is to measure spatial skill. If the forecast has no spatial skill (that is, if it is random), that rate equals the cloud fraction $f_{\mathrm{cloud}}$ on the same tiles. If the forecast is independent of the true sky, then $P(\text{pred clear} \cap \text{true cloud}) = P(\text{pred clear})\,P(\text{true cloud})$, and
\[
P(\text{true cloud} \mid \text{pred clear})
= \frac{P(\text{pred clear})\,P(\text{true cloud})}{P(\text{pred clear})}
= P(\text{true cloud})
= f_{\mathrm{cloud}}.
\]
Skill is a departure from the line $P(\text{true cloud} \mid \text{pred clear}) = f_{\mathrm{cloud}}$, and a smaller rate is a safer clear call.

\begin{figure}[htbp]
\centering
\includegraphics[width=\linewidth]{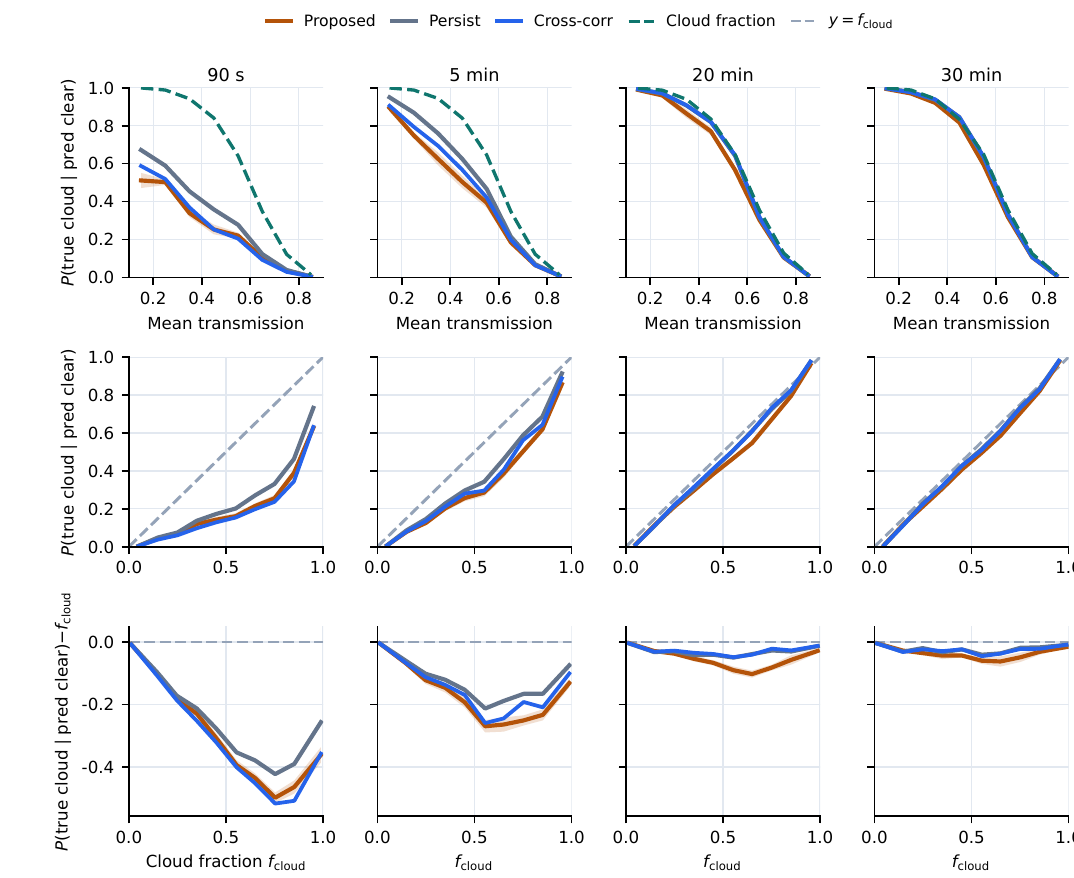}
\caption{Clear-call risk on the same holdout as Tab.~\ref{tab:mean_t_01_09}. The vertical axis is $P(\mathrm{true\ cloud}\mid\mathrm{pred\ clear})$: a tile called clear at forecast $T>0.63$ that is cloudy in the observed DREAM map at $T\le 0.63$. Top: that rate against the window-mean transmission. Middle: the rate against cloud fraction, with the dashed line $y=f_{\mathrm{cloud}}$. Bottom: the gap under that line. The shaded band on Proposed is the standard deviation across the four training seeds. Lower is a safer clear call. Columns are the four horizons.}
\label{fig:fc-vs-cover}
\end{figure}

Fig.~\ref{fig:fc-vs-cover} shows this rate on the $T\in[0.1,0.9]$ holdout. The rate rises with cloud fraction at every lead. The useful quantity is the gap under the diagonal. It is deepest on mixed skies, about $-0.50$ near $f_{\mathrm{cloud}}\sim 0.7$--$0.8$ at $90$\,s and about $-0.27$ near $f_{\mathrm{cloud}}\sim 0.5$--$0.6$ at $5$\,min. By $20$ and $30$\,min the deepest gap is only about $-0.10$ and $-0.06$, around $60$--$70\%$ cloud cover. Nearly clear skies sit on the line at every lead, because both the rate and $f_{\mathrm{cloud}}$ are already near zero. The overcast end does the same at $20$--$30$\,min. At $90$\,s an overcast clear call is still well under the line, so those edges are not a uniformly random forecast. At the pair-visit leads, cross-correlation has fallen onto persistence, and the proposed model remains slightly below both, indicating skill.

\section{Bulk crossing and condensing build-up}
\label{sect:app-buildup}

We are also interested in how clouds build up by two mechanisms, bulk crossing versus condensing, and how that affects performance. It has already been demonstrated that MAYE improves when the cloud fraction is changing in either direction (Fig.~\ref{fig:skill-regimes}). That score leaves out whether a clear patch is placed better than chance, and it mixes two ways a deck can build up. One is bulk motion across the field. The other is cloud that a single wind shift of the last frame does not explain. The second case is expected to be harder than bulk motion. 

\begin{figure}[htbp]
\centering
\includegraphics[width=0.92\linewidth]{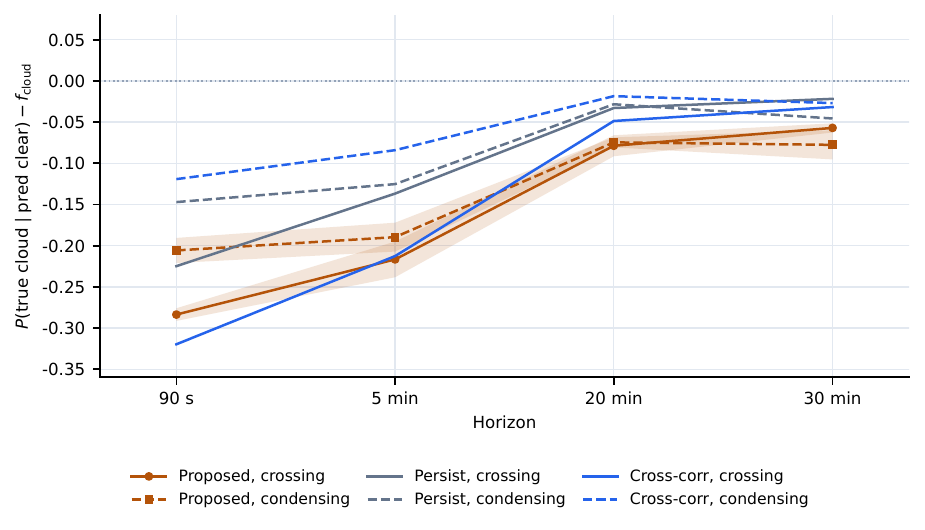}
\caption{Build-up only, on the same holdout. The vertical axis is $P(\mathrm{true\ cloud}\mid\mathrm{pred\ clear})-f_{\mathrm{cloud}}$, as in the bottom row of Fig.~\ref{fig:fc-vs-cover}. Solid lines: one wind shift of the last frame matches the future cloud better than leaving it in place. Dashed lines: it does not. The shaded band is the standard deviation across the four training seeds.}
\label{fig:fc-stories}
\end{figure}

In Fig.~\ref{fig:fc-stories} we isolate mixed skies, $f_{\mathrm{cloud}}\in[0.2,0.8]$, whose cloud fraction rose by at least $0.15$. A window is defined as \textit{crossing} when the cloud mask of one wind shift overlaps the future cloud more than the last frame left in place. It is defined as \textit{condensing} when the shift does no better.

At $90$\,s the proposed gap between the two phases is clear, with condensing worse than bulk motion. Cross-correlation splits them further. By $20$ and $30$\,min the two proposed curves sit together, about $0.06$--$0.08$ under the line, still about twice as far under it as persistence. The satellite forecast can tell that the field is filling in. It does not, however, keep the remaining clear tiles on the correct HEALPix cells, so the two build-up stories meet on this axis.

This is expected. The camera field above $30^\circ$ altitude is about $3.5$ times the cloud height above the summit. Cloud-top height is a proxy for that height. On one night within $25$\,km the median top is about $5.3$\,km, with the 10th--90th percentile at $3$--$10$\,km. The summit is at $2.6$\,km, so that percentile range sits about $1$--$7$\,km above the camera and the median about $2.7$\,km. The field above $30^\circ$ is then about $10$--$25$\,km across. The same decks on the full camera, down to $\sim$20$^\circ$, span about $5$--$40$\,km. The satellite imagery has a cadence of about 8--10\,min. At $15$--$20$\,m\,s$^{-1}$, one frame moves a cloud about $7$--$12$\,km, so a typical deck crosses the scored field in about $10$--$25$\,min, one or two scans. The lack of skill separation, even though the satellite can see beyond the horizon, is likely bottlenecked by that cadence and by the height of the incoming deck. 

\bibliographystyle{spiejour}
\bibliography{references}

@article{Ivezic2019,
  author  = "{\v{Z}}. Ivezi{\'c} and S. M. Kahn and J. A. Tyson and others",
  title   = "{LSST}: from science drivers to reference design and anticipated data products",
  journal = "Astrophys. J.",
  volume  = "873",
  pages   = "111",
  year    = "2019",
  doi     = "10.3847/1538-4357/ab042c",
  note    = "[doi:10.3847/1538-4357/ab042c]"
}

@article{Naghib2019,
  author  = "E. Naghib and P. Yoachim and R. J. Vanderbei and A. J. Connolly and R. L. Jones",
  title   = "A framework for telescope schedulers: with applications to the {Large Synoptic Survey Telescope}",
  journal = "Astron. J.",
  volume  = "157",
  pages   = "151",
  year    = "2019",
  doi     = "10.3847/1538-3881/aafece",
  note    = "[doi:10.3847/1538-3881/aafece]"
}

@inproceedings{Stuik2024,
  author    = "R. Stuik and S. Timmer and P. Vreeswijk and K. Fanning and H. Hoekstra and B. Stalder and P. Kub{\'a}nek and M. Logue and W. van Reeven and T. Ribeiro and P. Ingraham and S. Bloemen and K. Kuijken",
  title     = "All-sky transparency mapping for the {Vera C. Rubin Observatory} with {DREAM}",
  booktitle = "Ground-based and Airborne Instrumentation for Astronomy X",
  series    = "Proc. SPIE",
  volume    = "13096",
  pages     = "1309632",
  year      = "2024",
  doi       = "10.1117/12.3018302",
  note      = "[doi:10.1117/12.3018302]"
}

@inproceedings{Stuik2022,
  author    = "R. Stuik and S. T. Timmer and H. Hoekstra and P. Ingraham and W. van Reeven and T. Ribeiro and others",
  title     = "The {Dutch Rubin Enhanced Atmospheric Monitor} --- {DREAM}",
  booktitle = "Ground-based and Airborne Instrumentation for Astronomy IX",
  series    = "Proc. SPIE",
  volume    = "12184",
  year      = "2022",
  doi       = "10.1117/12.2629329",
  note      = "[doi:10.1117/12.2629329]"
}

@article{Talens2017,
  author  = "G. J. J. Talens and J. F. P. Spronck and A.-L. Lesage and G. P. P. L. Otten and R. Stuik and D. Pollacco and I. A. G. Snellen",
  title   = "The {Multi-site All-Sky CAmeRA} ({MASCARA}): finding transiting exoplanets around bright ($m_V < 8$) stars",
  journal = "Astron. Astrophys.",
  volume  = "601",
  pages   = "A11",
  year    = "2017",
  doi     = "10.1051/0004-6361/201630319",
  note    = "[doi:10.1051/0004-6361/201630319]"
}

@article{Talens2018,
  author  = "G. J. J. Talens and E. R. Deul and R. Stuik and others",
  title   = "Data calibration for the {MASCARA} and {bRing} instruments",
  journal = "Astron. Astrophys.",
  volume  = "619",
  pages   = "A154",
  year    = "2018",
  doi     = "10.1051/0004-6361/201834070",
  note    = "[doi:10.1051/0004-6361/201834070]"
}

@article{Gorski2005,
  author  = "K. M. G{\'o}rski and E. Hivon and A. J. Banday and B. D. Wandelt and F. K. Hansen and M. Reinecke and M. Bartelmann",
  title   = "{HEALPix}: a framework for high-resolution discretization and fast analysis of data distributed on the sphere",
  journal = "Astrophys. J.",
  volume  = "622",
  pages   = "759-771",
  year    = "2005",
  doi     = "10.1086/427976",
  note    = "[doi:10.1086/427976]"
}

@article{Schmit2017,
  author  = "T. J. Schmit and P. Griffith and M. M. Gunshor and J. M. Daniels and S. J. Goodman and W. J. Lebair",
  title   = "A closer look at the {ABI} on the {GOES-R} series",
  journal = "Bull. Am. Meteorol. Soc.",
  volume  = "98",
  pages   = "681-698",
  year    = "2017",
  doi     = "10.1175/BAMS-D-15-00213.1",
  note    = "[doi:10.1175/BAMS-D-15-00213.1]"
}

@article{Heidinger2009,
  author  = "A. K. Heidinger and M. J. Pavolonis",
  title   = "Gazing at Cirrus Clouds for 25 Years through a Split Window. {Part I}: Methodology",
  journal = "J. Appl. Meteorol. Climatol.",
  volume  = "48",
  pages   = "1100-1116",
  year    = "2009",
  doi     = "10.1175/2008JAMC1946.1",
  note    = "[doi:10.1175/2008JAMC1946.1]"
}

@article{Heidinger2012,
  author  = "A. K. Heidinger and A. T. Evan and M. J. Foster and A. Walther",
  title   = "A naive {Bayesian} cloud-detection scheme derived from {CALIPSO} and applied within {PATMOS-x}",
  journal = "J. Appl. Meteorol. Climatol.",
  volume  = "51",
  pages   = "1129-1144",
  year    = "2012",
  doi     = "10.1175/JAMC-D-11-02.1",
  note    = "[doi:10.1175/JAMC-D-11-02.1]"
}

@article{Bresky2012,
  author  = "W. C. Bresky and J. M. Daniels and A. A. Bailey and S. T. Wanzong",
  title   = "New methods toward minimizing the slow speed bias associated with atmospheric motion vectors",
  journal = "J. Appl. Meteorol. Climatol.",
  volume  = "51",
  pages   = "2137-2151",
  year    = "2012",
  doi     = "10.1175/JAMC-D-11-0234.1",
  note    = "[doi:10.1175/JAMC-D-11-0234.1]"
}

@inproceedings{Lucas1981,
  author    = "B. D. Lucas and T. Kanade",
  title     = "An iterative image registration technique with an application to stereo vision",
  booktitle = "Proc. 7th Int. Joint Conf. on Artificial Intelligence",
  pages     = "674-679",
  year      = "1981"
}

@article{Bowler2006,
  author  = "N. E. Bowler and C. E. Pierce and A. Seed",
  title   = "Development of a precipitation nowcasting algorithm based upon optical flow techniques",
  journal = "J. Hydrol.",
  volume  = "328",
  pages   = "74-83",
  year    = "2006",
  doi     = "10.1016/j.jhydrol.2005.11.049",
  note    = "[doi:10.1016/j.jhydrol.2005.11.049]"
}

@article{Seed2003,
  author  = "A. W. Seed",
  title   = "A dynamic and spatial scaling approach to advection forecasting",
  journal = "J. Appl. Meteorol.",
  volume  = "42",
  pages   = "381-388",
  year    = "2003",
  doi     = "10.1175/1520-0450(2003)042<0381:ADASSA>2.0.CO;2",
  note    = "[doi:10.1175/1520-0450(2003)042<0381:ADASSA>2.0.CO;2]"
}

@article{Pulkkinen2019,
  author  = "S. Pulkkinen and D. Nerini and A. A. P{\'e}rez Hortal and C. Velasco-Forero and A. Seed and U. Germann and L. Foresti",
  title   = "{Pysteps}: an open-source {Python} library for probabilistic precipitation nowcasting ({v1.0})",
  journal = "Geosci. Model Dev.",
  volume  = "12",
  pages   = "4185-4219",
  year    = "2019",
  doi     = "10.5194/gmd-12-4185-2019",
  note    = "[doi:10.5194/gmd-12-4185-2019]"
}

@article{Chow2011,
  author  = "C. W. Chow and B. Urquhart and M. Lave and A. Dominguez and J. Kleissl and J. Shields and B. Washom",
  title   = "Intra-hour forecasting with a total sky imager at the {UC San Diego} solar energy testbed",
  journal = "Sol. Energy",
  volume  = "85",
  pages   = "2881-2893",
  year    = "2011",
  doi     = "10.1016/j.solener.2011.08.025",
  note    = "[doi:10.1016/j.solener.2011.08.025]"
}

@article{Jones2018,
  author  = "R. L. Jones and P. J. Brown and P. A. Abell and others",
  title   = "The {Large Synoptic Survey Telescope} as a {Near-Earth Object} discovery machine",
  journal = "Icarus",
  volume  = "303",
  pages   = "181-202",
  year    = "2018",
  doi     = "10.1016/j.icarus.2017.11.033",
  note    = "[doi:10.1016/j.icarus.2017.11.033]"
}

@article{Sanchez2022,
  author  = "B. O. S{\'a}nchez and others",
  title   = "{SNIa} Cosmology Analysis Results from Simulated {LSST} Images: From Difference Imaging to Constraints on Dark Energy",
  journal = "Astrophys. J.",
  volume  = "934",
  pages   = "96",
  year    = "2022",
  doi     = "10.3847/1538-4357/ac7a37",
  note    = "[doi:10.3847/1538-4357/ac7a37]"
}

@techreport{LSST_OSS,
  author      = "C. Claver and others",
  title       = "{LSST} Observatory System Specifications",
  institution = "Vera C. Rubin Observatory",
  number      = "LSE-30",
  year        = "2021"
}

@article{Bianco2022,
  author  = "F. B. Bianco and {\v{Z}}. Ivezi{\'c} and R. L. Jones and others",
  title   = "Optimization of the Observing Cadence for the {Rubin Observatory Legacy Survey of Space and Time}: a pioneering process of community-focused experimental design",
  journal = "Astrophys. J. Suppl. Ser.",
  volume  = "258",
  pages   = "1",
  year    = "2022",
  doi     = "10.3847/1538-4365/ac3e72",
  note    = "[doi:10.3847/1538-4365/ac3e72]"
}

@techreport{Jones2021,
  author      = "R. L. Jones and P. Yoachim and {\v{Z}}. Ivezi{\'c} and E. H. Neilsen, Jr. and T. Ribeiro",
  title       = "Survey Strategy and Cadence Choices for the {Vera C. Rubin Observatory Legacy Survey of Space and Time} ({LSST})",
  institution = "Vera C. Rubin Observatory",
  number      = "PSTN-051",
  year        = "2021",
  doi         = "10.71929/rubin/2584084",
  note        = "[doi:10.71929/rubin/2584084]"
}

@techreport{PSTN055,
  author      = "F. E. Bauer and F. B. Bianco and S. Brough and others",
  title       = "Survey Cadence Optimization Committee's Phase 2 Recommendations",
  institution = "Vera C. Rubin Observatory",
  number      = "PSTN-055",
  year        = "2023",
  doi         = "10.71929/rubin/2585249",
  note        = "[doi:10.71929/rubin/2585249]"
}

@misc{LSSTObservingStrategy,
  author       = "{Vera C. Rubin Observatory}",
  title        = "{LSST} Baseline Strategy --- Observing Strategy",
  year         = "2026",
  howpublished = "https://survey-strategy.lsst.io/",
  note         = "Baseline v5.3.0; WFD cadence last updated 28 June 2026. WFD uses slightly more than 80 percent of survey time; visits are typically mixed-filter pairs at about 33 min; about 3 percent of pairs (4 percent of visits) receive a later third visit."
}

@article{Schwamb2023,
  author  = "M. E. Schwamb and others",
  title   = "Tuning the {Legacy Survey of Space and Time} ({LSST}) Observing Strategy for Solar System Science",
  journal = "Astrophys. J. Suppl. Ser.",
  volume  = "266",
  pages   = "22",
  year    = "2023",
  doi     = "10.3847/1538-4365/acc173",
  note    = "[doi:10.3847/1538-4365/acc173]"
}

@inproceedings{Yoachim2016,
  author    = "P. Yoachim and M. W. Coughlin and G. Z. Angeli and others",
  title     = "An optical to {IR} sky brightness model for the {LSST}",
  booktitle = "Observatory Operations: Strategies, Processes, and Systems VI",
  series    = "Proc. SPIE",
  volume    = "9910",
  pages     = "99101A",
  year      = "2016",
  doi       = "10.1117/12.2232947",
  note      = "[doi:10.1117/12.2232947]"
}

@techreport{SMTN002,
  author      = "R. L. Jones",
  title       = "Calculating {LSST} limiting magnitudes and {SNR}",
  institution = "Vera C. Rubin Observatory",
  number      = "SMTN-002",
  year        = "2017"
}

@techreport{TSTN035,
  author      = "T. Ribeiro and others",
  title       = "Handling Targets of Opportunity",
  institution = "Vera C. Rubin Observatory",
  number      = "TSTN-035",
  year        = "2023"
}

@article{Germann2002,
  author  = "U. Germann and I. Zawadzki",
  title   = "Scale-dependence of the predictability of precipitation from continental radar images. {Part I}: Description of the methodology",
  journal = "Mon. Weather Rev.",
  volume  = "130",
  pages   = "2859-2873",
  year    = "2002",
  doi     = "10.1175/1520-0493(2002)130<2859:SDOTPO>2.0.CO;2",
  note    = "[doi:10.1175/1520-0493(2002)130<2859:SDOTPO>2.0.CO;2]"
}

@article{Staniforth1991,
  author  = "A. Staniforth and J. C{\^o}t{\'e}",
  title   = "Semi-{Lagrangian} integration schemes for atmospheric models---a review",
  journal = "Mon. Weather Rev.",
  volume  = "119",
  pages   = "2206-2223",
  year    = "1991",
  doi     = "10.1175/1520-0493(1991)119<2206:SLISFA>2.0.CO;2",
  note    = "[doi:10.1175/1520-0493(1991)119<2206:SLISFA>2.0.CO;2]"
}

@techreport{Bouguet2000,
  author      = "J.-Y. Bouguet",
  title       = "Pyramidal implementation of the {Lucas} {Kanade} feature tracker: description of the algorithm",
  institution = "Intel Corporation, Microprocessor Research Labs",
  year        = "2000"
}

@article{Yang2019,
  author  = "D. Yang",
  title   = "A universal benchmarking method for probabilistic solar irradiance forecasting",
  journal = "Sol. Energy",
  volume  = "184",
  pages   = "410-416",
  year    = "2019",
  doi     = "10.1016/j.solener.2019.04.018",
  note    = "[doi:10.1016/j.solener.2019.04.018]"
}

@techreport{Heidinger2020,
  author      = "A. Heidinger and Y. Li and S. Wanzong",
  title       = "Enterprise {AWG} Cloud Height Algorithm ({ACHA})",
  institution = "NOAA/NESDIS/STAR",
  number      = "Algorithm Theoretical Basis Document, Version 3.4",
  year        = "2020",
  note        = "\url{https://www.star.nesdis.noaa.gov/goesr/documents/ATBDs/Enterprise/ATBD_Enterprise_Cloud_Height_v3.4_2020-09.pdf}"
}

@techreport{Daniels2019,
  author      = "J. Daniels and W. Bresky and S. Wanzong and C. Velden and H. Berger",
  title       = "{GOES-R} Advanced Baseline Imager ({ABI}) Algorithm Theoretical Basis Document for Derived Motion Winds",
  institution = "NOAA/NESDIS/STAR",
  number      = "Version 3.1",
  year        = "2019",
  note        = "\url{https://www.star.nesdis.noaa.gov/goesr/documents/ATBDs/Baseline/ATBD_GOES-R_Winds_v3.1_Feb2019.pdf}"
}

@article{Velden2005,
  author  = "C. Velden and J. Daniels and D. Stettner and D. Santek and J. Key and J. Dunion and K. Holmlund and G. Dengel and W. Bresky and P. Menzel",
  title   = "Recent innovations in deriving tropospheric winds from meteorological satellites",
  journal = "Bull. Am. Meteorol. Soc.",
  volume  = "86",
  pages   = "205-223",
  year    = "2005",
  doi     = "10.1175/BAMS-86-2-205",
  note    = "[doi:10.1175/BAMS-86-2-205]"
}

@misc{NOAAGoes19,
  author       = "{NOAA Office of Satellite and Product Operations}",
  title        = "{GOES}-19 operational declaration as {GOES}-{East}",
  year         = "2025",
  howpublished = "ESPC Notification MSG\_20250407\_1510, 7 April 2025, 1510 UTC",
  note         = "\url{https://www.ospo.noaa.gov/data/messages/2025/04/MSG_20250407_1510.html}"
}

@phdthesis{Segev2026,
  author  = "L. Segev",
  title   = "Cloudy with a Chance of Data: A Dynamic Scheduler for All-Sky Surveys with Real-Time Optimization",
  school  = "Harvard University, Department of Astronomy",
  year    = "2026",
  note    = "BA thesis; advisor C.\ W.\ Stubbs"
}

@misc{Benn1998,
  author       = "C. R. Benn and W. Martin",
  title        = "Observing overheads at the {William Herschel Telescope}",
  year         = "1998",
  howpublished = "ING Technical Note; analysis of WHT observing logs",
  note         = "Bad weather accounted for 24 percent of scheduled observing time; http://www.ing.iac.es/Astronomy/telescopes/wht/overheads.html"
}

@article{Qian2024,
  author  = "X. Qian and Y. Yao and H. Wang and others",
  title   = "Statistics of cloud cover above the {Ali Observatory}, {Tibet}",
  journal = "Mon. Not. R. Astron. Soc.",
  volume  = "529",
  pages   = "1195-1204",
  year    = "2024",
  doi     = "10.1093/mnras/stae582",
  note    = "[doi:10.1093/mnras/stae582]"
}

@article{Xu2023,
  author  = "J. Xu and G.-j. Feng and G.-x. Pu and others",
  title   = "Site-testing at the {Muztagh-ata} Site {V}. Nighttime Cloud Amount during the Last Five Years",
  journal = "Res. Astron. Astrophys.",
  volume  = "23",
  pages   = "045015",
  year    = "2023",
  doi     = "10.1088/1674-4527/acc29b",
  note    = "[doi:10.1088/1674-4527/acc29b]"
}

@article{Patat2011,
  author  = "F. Patat and S. Moehler and K. O'Brien and others",
  title   = "Optical atmospheric extinction over {Cerro Paranal}",
  journal = "Astron. Astrophys.",
  volume  = "527",
  pages   = "A91",
  year    = "2011",
  doi     = "10.1051/0004-6361/201015537",
  note    = "[doi:10.1051/0004-6361/201015537]"
}

@inproceedings{Kerber2014,
  author    = "F. Kerber and R. R. Querel and R. Hanuschik",
  title     = "Quantifying photometric observing conditions on {Paranal} using an {IR} camera",
  booktitle = "Observatory Operations: Strategies, Processes, and Systems V",
  series    = "Proc. SPIE",
  volume    = "9149",
  pages     = "91490M",
  year      = "2014",
  doi       = "10.1117/12.2055168",
  note      = "[doi:10.1117/12.2055168]"
}

@inproceedings{Kerber2016,
  author    = "F. Kerber and R. R. Querel and B. Neureiter and R. Hanuschik",
  title     = "Through thick and thin: quantitative classification of photometric observing conditions on {Paranal}",
  booktitle = "Observatory Operations: Strategies, Processes, and Systems VI",
  series    = "Proc. SPIE",
  volume    = "9910",
  pages     = "99101S",
  year      = "2016",
  doi       = "10.1117/12.2232333",
  note      = "[doi:10.1117/12.2232333]"
}

@article{Haslebacher2022,
  author  = "C. Haslebacher and M.-E. Demory and B.-O. Demory and M. Sarazin and P. L. Vidale",
  title   = "Impact of climate change on site characteristics of eight major astronomical observatories using high-resolution global climate projections until 2050",
  journal = "Astron. Astrophys.",
  volume  = "665",
  pages   = "A149",
  year    = "2022",
  doi     = "10.1051/0004-6361/202142493",
  note    = "[doi:10.1051/0004-6361/202142493]"
}

@article{Snellen2013,
  author  = "I. Snellen and R. Stuik and G. Otten and others",
  title   = "{MASCARA}: The Multi-site All-Sky {CAmeRA}",
  journal = "EPJ Web Conf.",
  volume  = "47",
  pages   = "03008",
  year    = "2013",
  doi     = "10.1051/epjconf/20134703008",
  note    = "[doi:10.1051/epjconf/20134703008]"
}

@article{Stuik2017,
  author  = "R. Stuik and J. I. Bailey and P. Dorval and others",
  title   = "{bRing}: An observatory dedicated to monitoring the {$\beta$ Pictoris} b Hill sphere transit",
  journal = "Astron. Astrophys.",
  volume  = "607",
  pages   = "A45",
  year    = "2017",
  doi     = "10.1051/0004-6361/201731679",
  note    = "[doi:10.1051/0004-6361/201731679]"
}

@article{Lochner2022,
  author  = "M. Lochner and D. Scolnic and H. Almoubayyed and others",
  title   = "The Impact of Observing Strategy on Cosmological Constraints with {LSST}",
  journal = "Astrophys. J. Suppl. Ser.",
  volume  = "259",
  pages   = "58",
  year    = "2022",
  doi     = "10.3847/1538-4365/ac5033",
  note    = "[doi:10.3847/1538-4365/ac5033]"
}

@article{Tokovinin2023,
  author  = "A. Tokovinin",
  title   = "The Elusive Nature of ``Seeing''",
  journal = "Atmosphere",
  volume  = "14",
  pages   = "1694",
  year    = "2023",
  doi     = "10.3390/atmos14111694",
  note    = "[doi:10.3390/atmos14111694]"
}

@article{Masciadri2013,
  author  = "E. Masciadri and F. Lascaux and L. Fini",
  title   = "{MOSE}: operational forecast of the optical turbulence and atmospheric parameters at {European Southern Observatory} ground-based sites -- {I}. Overview and vertical stratification of atmospheric parameters at 0--20\,km",
  journal = "Mon. Not. R. Astron. Soc.",
  volume  = "436",
  pages   = "1968-1985",
  year    = "2013",
  doi     = "10.1093/mnras/stt1708",
  note    = "[doi:10.1093/mnras/stt1708]"
}

@inproceedings{Shi2015,
  author    = "X. Shi and Z. Chen and H. Wang and D.-Y. Yeung and W.-K. Wong and W.-C. Woo",
  title     = "Convolutional {LSTM} network: a machine learning approach for precipitation nowcasting",
  booktitle = "Adv. Neural Inf. Process. Syst.",
  volume    = "28",
  year      = "2015"
}

@inproceedings{Shi2017,
  author    = "X. Shi and Z. Gao and L. Lausen and H. Wang and D.-Y. Yeung and W.-K. Wong and W.-C. Woo",
  title     = "Deep learning for precipitation nowcasting: a benchmark and a new model",
  booktitle = "Adv. Neural Inf. Process. Syst.",
  volume    = "30",
  year      = "2017"
}

@inproceedings{Wang2017,
  author    = "Y. Wang and M. Long and J. Wang and Z. Gao and P. S. Yu",
  title     = "{PredRNN}: recurrent neural networks for predictive learning using spatiotemporal {LSTMs}",
  booktitle = "Adv. Neural Inf. Process. Syst.",
  volume    = "30",
  year      = "2017"
}

@inproceedings{LeGuen2020,
  author    = "V. Le Guen and N. Thome",
  title     = "Disentangling physical dynamics from unknown factors for unsupervised video prediction",
  booktitle = "Proc. {IEEE}/{CVF} Conf. Comput. Vis. Pattern Recognit.",
  pages     = "11474-11484",
  year      = "2020",
  doi       = "10.1109/CVPR42600.2020.01149",
  note      = "[doi:10.1109/CVPR42600.2020.01149]"
}

@inproceedings{Gao2022,
  author    = "Z. Gao and C. Tan and L. Wu and S. Z. Li",
  title     = "{SimVP}: simpler yet better video prediction",
  booktitle = "Proc. {IEEE}/{CVF} Conf. Comput. Vis. Pattern Recognit.",
  pages     = "3170-3180",
  year      = "2022",
  doi       = "10.1109/CVPR52688.2022.00317",
  note      = "[doi:10.1109/CVPR52688.2022.00317]"
}

@inproceedings{Gao2022Earthformer,
  author    = "Z. Gao and X. Shi and H. Wang and Y. Zhu and Y. B. Wang and M. Li and D.-Y. Yeung",
  title     = "{Earthformer}: exploring space-time transformers for earth system forecasting",
  booktitle = "Adv. Neural Inf. Process. Syst.",
  volume    = "35",
  year      = "2022"
}

@misc{Sonderby2020,
  author       = "C. K. S{\o}nderby and L. Espeholt and J. Heek and M. Dehghani and A. Oliver and T. Salimans and S. Agrawal and J. Hickey and N. Kalchbrenner",
  title        = "{MetNet}: a neural weather model for precipitation forecasting",
  year         = "2020",
  howpublished = "arXiv:2003.12140"
}

@article{Ravuri2021,
  author  = "S. Ravuri and K. Lenc and M. Willson and D. Kangin and R. Lam and P. Mirowski and M. Fitzsimons and M. Athanassiadou and S. Kashem and S. Madge and others",
  title   = "Skilful precipitation nowcasting using deep generative models of radar",
  journal = "Nature",
  volume  = "597",
  pages   = "672-677",
  year    = "2021",
  doi     = "10.1038/s41586-021-03854-z",
  note    = "[doi:10.1038/s41586-021-03854-z]"
}

@article{Zhang2023,
  author  = "Y. Zhang and M. Long and K. Chen and X. Xing and J. Wang and M. Li and W. Yin and X. Hu and W. Ouyang and J. Wang and others",
  title   = "Skilful nowcasting of extreme precipitation with {NowcastNet}",
  journal = "Nature",
  volume  = "619",
  pages   = "526-532",
  year    = "2023",
  doi     = "10.1038/s41586-023-06184-4",
  note    = "[doi:10.1038/s41586-023-06184-4]"
}

\end{spacing}
\end{document}